\documentclass[aps,prd,twocolumn,superscriptaddress,nofootinbib]{revtex4}

\usepackage{graphicx}
\usepackage{amsmath}
\usepackage{dcolumn}

\begin{document}


\title{Search for Gravitational Waves Associated with 39 Gamma-Ray Bursts\\
       Using Data from the Second, Third, and Fourth LIGO Runs\\
{\small LIGO-P060024-07-Z}}

\affiliation{Albert-Einstein-Institut, Max-Planck-Institut f\"ur Gravitationsphysik, D-14476 Golm, Germany}
\affiliation{Albert-Einstein-Institut, Max-Planck-Institut f\"ur Gravitationsphysik, D-30167 Hannover, Germany}
\affiliation{Andrews University, Berrien Springs, MI 49104 USA}
\affiliation{Australian National University, Canberra, 0200, Australia}
\affiliation{California Institute of Technology, Pasadena, CA  91125, USA}
\affiliation{California State University Dominguez Hills, Carson, CA  90747, USA}
\affiliation{Caltech-CaRT, Pasadena, CA  91125, USA}
\affiliation{Cardiff University, Cardiff, CF24 3AA, United Kingdom}
\affiliation{Carleton College, Northfield, MN  55057, USA}
\affiliation{Charles Sturt University, Wagga Wagga, NSW 2678, Australia}
\affiliation{Columbia University, New York, NY  10027, USA}
\affiliation{Embry-Riddle Aeronautical University, Prescott, AZ   86301 USA}
\affiliation{Hobart and William Smith Colleges, Geneva, NY  14456, USA}
\affiliation{Inter-University Centre for Astronomy  and Astrophysics, Pune - 411007, India}
\affiliation{LIGO - California Institute of Technology, Pasadena, CA  91125, USA}
\affiliation{LIGO Hanford Observatory, Richland, WA  99352, USA}
\affiliation{LIGO Livingston Observatory, Livingston, LA  70754, USA}
\affiliation{LIGO - Massachusetts Institute of Technology, Cambridge, MA 02139, USA}
\affiliation{Louisiana State University, Baton Rouge, LA  70803, USA}
\affiliation{Louisiana Tech University, Ruston, LA  71272, USA}
\affiliation{Loyola University, New Orleans, LA 70118, USA}
\affiliation{Max Planck Institut f\"ur Quantenoptik, D-85748, Garching, Germany}
\affiliation{Moscow State University, Moscow, 119992, Russia}
\affiliation{NASA/Goddard Space Flight Center, Greenbelt, MD  20771, USA}
\affiliation{National Astronomical Observatory of Japan, Tokyo  181-8588, Japan}
\affiliation{Northwestern University, Evanston, IL  60208, USA}
\affiliation{Rochester Institute of Technology, Rochester, NY 14623, USA}
\affiliation{Rutherford Appleton Laboratory, Chilton, Didcot, Oxon OX11 0QX United Kingdom}
\affiliation{San Jose State University, San Jose, CA 95192, USA}
\affiliation{Southeastern Louisiana University, Hammond, LA  70402, USA}
\affiliation{Southern University and A\&M College, Baton Rouge, LA  70813, USA}
\affiliation{Stanford University, Stanford, CA  94305, USA}
\affiliation{Syracuse University, Syracuse, NY  13244, USA}
\affiliation{The Pennsylvania State University, University Park, PA  16802, USA}
\affiliation{The University of Texas at Brownsville and Texas Southmost College, Brownsville, TX  78520, USA}
\affiliation{Trinity University, San Antonio, TX  78212, USA}
\affiliation{Universitat de les Illes Balears, E-07122 Palma de Mallorca, Spain}
\affiliation{Universit\"at Hannover, D-30167 Hannover, Germany}
\affiliation{University of Adelaide, Adelaide, SA 5005, Australia}
\affiliation{University of Birmingham, Birmingham, B15 2TT, United Kingdom}
\affiliation{University of Florida, Gainesville, FL  32611, USA}
\affiliation{University of Glasgow, Glasgow, G12 8QQ, United Kingdom}
\affiliation{University of Maryland, College Park, MD 20742 USA}
\affiliation{University of Michigan, Ann Arbor, MI  48109, USA}
\affiliation{University of Oregon, Eugene, OR  97403, USA}
\affiliation{University of Rochester, Rochester, NY  14627, USA}
\affiliation{University of Salerno, 84084 Fisciano (Salerno), Italy}
\affiliation{University of Sannio at Benevento, I-82100 Benevento, Italy}
\affiliation{University of Sheffield, Sheffield, S3 7RH, United Kingdom}
\affiliation{University of Southampton, Southampton, SO17 1BJ, United Kingdom}
\affiliation{University of Strathclyde, Glasgow, G1 1XQ, United Kingdom}
\affiliation{University of Washington, Seattle, WA, 98195}
\affiliation{University of Western Australia, Crawley, WA 6009, Australia}
\affiliation{University of Wisconsin-Milwaukee, Milwaukee, WI  53201, USA}
\affiliation{Vassar College, Poughkeepsie, NY 12604}
\affiliation{Washington State University, Pullman, WA 99164, USA}
\author{B.~Abbott}\affiliation{LIGO - California Institute of Technology, Pasadena, CA  91125, USA}
\author{R.~Abbott}\affiliation{LIGO - California Institute of Technology, Pasadena, CA  91125, USA}
\author{R.~Adhikari}\affiliation{LIGO - California Institute of Technology, Pasadena, CA  91125, USA}
\author{J.~Agresti}\affiliation{LIGO - California Institute of Technology, Pasadena, CA  91125, USA}
\author{P.~Ajith}\affiliation{Albert-Einstein-Institut, Max-Planck-Institut f\"ur Gravitationsphysik, D-30167 Hannover, Germany}
\author{B.~Allen}\affiliation{Albert-Einstein-Institut, Max-Planck-Institut f\"ur Gravitationsphysik, D-30167 Hannover, Germany}\affiliation{University of Wisconsin-Milwaukee, Milwaukee, WI  53201, USA}
\author{R.~Amin}\affiliation{Louisiana State University, Baton Rouge, LA  70803, USA}
\author{S.~B.~Anderson}\affiliation{LIGO - California Institute of Technology, Pasadena, CA  91125, USA}
\author{W.~G.~Anderson}\affiliation{University of Wisconsin-Milwaukee, Milwaukee, WI  53201, USA}
\author{M.~Arain}\affiliation{University of Florida, Gainesville, FL  32611, USA}
% ...
\collaboration{The LIGO Scientific Collaboration, http://www.ligo.org}
\noaffiliation

\date{\today}

\begin{abstract}
We present the results of a search for short-duration gravitational-wave bursts
associated with 39 gamma-ray bursts (GRBs) detected by gamma-ray satellite
experiments during LIGO's S2, S3, and S4 science runs.  The search involves
calculating the crosscorrelation between two interferometer data streams
surrounding the GRB trigger time.  We search for associated gravitational
radiation from single GRBs, and also apply statistical tests to search for a
gravitational-wave signature associated with the whole sample.  For the sample
examined, we find no evidence for the association of gravitational radiation 
with GRBs, either on a single-GRB basis or on a statistical basis.  Simulating
gravitational-wave bursts with sine-gaussian waveforms, we set upper limits on 
the root-sum-square of the gravitational-wave strain amplitude of such waveforms
at the times of the GRB triggers.  We also demonstrate how a sample of several 
GRBs can be used collectively to set constraints on population models.  The 
small number of GRBs and the significant change in sensitivity of the detectors 
over the three runs, however, limits the usefulness of a population study for 
the S2, S3, and S4 runs.  Finally, we discuss prospects for the search 
sensitivity for the ongoing S5 run, and beyond for the next generation of 
detectors.
\end{abstract}


\maketitle


\section{Introduction}

It has been over three decades since gamma-ray bursts (GRBs) were first detected
by the {\it Vela} satellites \cite{klebesadel73}.  During the 1990s, when the
{\it Burst and Transient Source Experiment} (BATSE) \cite{fishman92} and BeppoSAX
\cite{boella97} were in operation, important discoveries and observations
relating to GRBs were made, such as the isotropic distribution of GRBs 
\cite{meegan92}; the bimodal distribution of burst durations, suggesting long 
and short classes of GRBs \cite{ck93}; detections of the first x-ray 
\cite{costa97}, optical \cite{paradijs97}, and radio \cite{frail97} 
counterparts; the first redshift measurements \cite{metzger97,cohen97,bloom98}; 
and the first hints of the association of long-duration GRBs with core-collapse 
supernovae \cite{galama98,iwamoto98,kulkarni98}.  Today, important questions 
about GRB progenitors, emission mechanisms and geometry linger, and observations 
made by the current generation of gamma-ray satellite experiments such as Swift 
\cite{swift04}, HETE-2 \cite{hete03}, INTEGRAL \cite{integral03}, and others
continue to provide new and exciting information which help us answer these
questions and better understand the origin and physics of these astrophysical
objects.

Currently favored models of GRB progenitors are core-collapse supernovae for
long-duration GRBs \cite{woosley93}, and neutron star-neutron star (NS-NS) or
neutron star-black hole (NS-BH) mergers for short-duration GRBs
\cite{schramm89,paczynski91}.  These models and the division into two classes of
progenitors are supported by observations of supernovae associated with
long-duration GRBs \cite{hjorth03,galama98,iwamoto98,kulkarni98,woosley99} and,
more recently, observations of afterglows and identification of host galaxies 
for short-duration GRBs \cite{gehrels05,villasenor05,fox05,hjorth05}.  The end 
result in either scenario is the formation of a stellar-mass black hole 
\cite{fryer99} and, in either scenario, theory predicts the emission of 
gravitational radiation.  In the former case, gravitational waves would result 
from the collapse of a massive star's core, while in the latter case, 
gravitational radiation would result from the inspiral, merger, and ringdown 
phases of the coalescence.  Recently, there has been an observation-driven 
suggestion of a third class of GRBs which could include both short- and 
long-duration GRBs \cite{grbclass3}, but more observations are needed to support 
this suggestion.

Due to the expected evolution of the proposed progenitors, the redshift 
distribution of long-duration GRBs is thought to follow the star formation rate
of the Universe \cite{jakobsson04,christensen04}, and recent redshift measurements
tend to support this model, with the measured GRB redshift distribution peaking
at $z\gtrsim1$ \cite{berger05}.  Long-duration GRBs have also been associated
exclusively with late-type star-forming host galaxies \cite{conselice05}.  On
the other hand, the recent observations of x-ray and optical afterglows from a
few short-duration bursts seem to suggest that these GRBs are located at lower
redshifts relative to long-duration GRBs \cite{guetta05,fox05}, and that short
bursts are found in a mixture of galaxy types, including elliptical galaxies,
which have older stellar populations.  All of these observations are consistent
with the currently favored models of GRB progenitors.  Although a large fraction
of GRBs are too distant for any associated GW signals to be detected by LIGO, it 
is plausible that a small fraction occur at closer distances.  For example,
a redshift of $z = 0.0085$, or a distance of 35 Mpc, has been associated with
long-duration burst/supernova GRB 980425/SN 1998bw \cite{galama98}.  It is not
unreasonable to expect that a few GRBs with no measured redshifts could have
been located relatively nearby as well.  For short-duration GRBs, the recent 
redshift observations have led to fairly optimistic estimates \cite{NYF,guetta06} 
for an associated GW observation in an extended LIGO science run.

In this paper, we present the results of a search for short-duration
gravitational-wave bursts (GWBs) associated with 39 GRBs that were detected by
gamma-ray satellite experiments on dates when the S2, S3, and S4 science
runs of the {\it Laser Interferometer Gravitational-Wave Observatory} (LIGO)
were in progress.  Although the theoretical shapes of the GW burst signals resulting 
from the two progenitor scenarios are not known, many models predict that the GW 
signals would be of short duration, ranging from $\sim$~1~ms to $\sim$~100~ms 
\cite{ottburrows04,dfm02,flanagan98,BHBHmerge1,BHBHmerge2}.  The search method 
presented here targets such short-duration signals, and calculates the 
crosscorrelation between two LIGO interferometer data streams to look for these 
signals.  A crosscorrelation-based method efficiently suppresses uncorrelated 
transient noise in the data streams, and at the same time tests that a candidate 
GW signal appears in data from at least two interferometers \cite{cadonati04}.
Previously, we presented the results of a search for a GWB associated with the 
bright and nearby GRB~030329 \cite{abbottgrb05}.  Here, we present analysis 
methods which search for GWBs associated with GRBs not only on an individual-GRB 
basis to target loud GWBs, but also on a statistical basis.  The statistical 
approach is sensitive to the cumulative effect of any weak GW signals that may be 
present in the LIGO data.

It is noted here that for the compact binary coalescence models of short-duration 
GRBs, a subset of the associated inspiral waveforms are well modelled, and that 
a template-based search for inspiral GW signals associated with short-duration 
GRBs is currently being developed using LIGO data.

\section{LIGO S2, S3, and S4 science runs}

The LIGO interferometers (IFOs) have been described in detail elsewhere
\cite{abbottnim04}.  These detectors are kilometer-length Michelson 
interferometers with orthogonal Fabry-Perot arms, designed to detect impinging
gravitational waves with frequencies ranging from $\sim40$~Hz to several kilohertz.
The interferometers' maximum sensitivity occurs near 100 Hz to 200 Hz.  There 
are two LIGO observatories: one located at Hanford, WA (LHO) and the other at 
Livingston, LA (LLO).  There are two IFOs at LHO:  one IFO with 4-km arms (H1) 
and the other with 2-km arms (H2).  The LLO observatory has one 4-km IFO (L1).  
The observatories are separated by a distance of 3000~km, corresponding to a
time-of-flight separation of 10~ms.

Each IFO consists of mirrors at the ends of each arm which serve as test masses.
Data from each IFO is in the form of a time series, digitized at 
16384~samples/s, which records the differential length of the arms and which, 
when calibrated, measures the strain induced by a gravitational-wave.  The 
response of an IFO to a given strain is measured by injecting sinusoidal 
excitations with known amplitude into the test mass control systems and tracking 
the resulting signals at the measurement point throughout each run.  The result 
is a measurement of the time-varying, frequency-dependent response function of 
each IFO.

\begin{figure}
\includegraphics[width=3.4in]{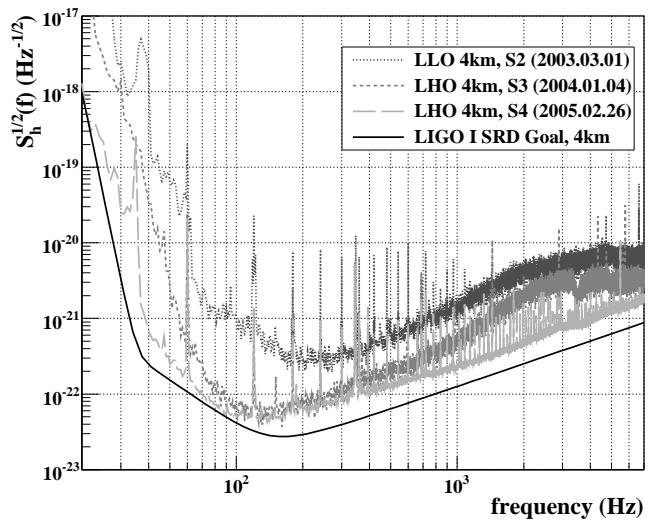}
\caption{\label{fig:ligoruns}Progression of LIGO sensitivities from S2 to S4
science runs.  For each run, the corresponding curve gives the magnitude of the 
noise spectral density, in strain-equivalent units, for one of the IFOs during a 
representative time interval within the run.  The solid curve gives the initial 
LIGO design sensitivity goal as given in LIGO's Science Requirements Document (SRD).}
\end{figure}

The LIGO S2 run was held from February to April 2003 (59 days), the S3 run from
October 2003 to January 2004 (70 days), and the S4 run from February to March
2005 (29 days).  The sensitivity of the LIGO detectors improved significantly
between the S2 and S4 runs, and approached the initial LIGO design sensitivity
during the LIGO S4 run.  The progression of the best LIGO sensitivity from the
S2 to S4 runs is shown in Fig.~\ref{fig:ligoruns}.  For each run, the 
corresponding curve in this plot gives the magnitude of the noise spectral 
density, in strain-equivalent units, for one of the IFOs during a representative
time interval within the run.  The solid curve gives the initial LIGO design 
sensitivity goal as given in LIGO's Science Requirements Document.  Further, the duty 
factor of the three IFOs increased significantly from the S2 to S4 run.  During 
the S2 run, the duty factors were 74\%, 58\%, and 37\% for the H1, H2, and L1 
IFOs, respectively, while during the S4 run, the duty factors were 
80.5\%, 81.4\%, and 74.5\%, respectively.

\section{The GRB sample}

Compared to the 1990s, when BATSE was detecting GRBs, the period from 2001 to 
2004, when LIGO had its first three science runs, was a time of relatively low 
GRB detection rate.  LIGO's S4 run coincided with a time when Swift had just 
started operating and was making its first GRB detections.  There were 29 GRB 
triggers during the S2 run, 11 GRB triggers during S3, and 6 GRB triggers during 
S4.  These GRB triggers were provided by the {\it Third Inter-Planetary Network} 
(IPN) \cite{ipn02}, Konus-Wind \cite{konus95}, HETE-2 , INTEGRAL, and Swift, and 
were distributed via the {\it GRB Coordinates Network}
(GCN).\footnote{http://gcn.gsfc.nasa.gov}

Only LIGO data which were of science mode quality were analyzed.  These science
mode segments are data collected when the interferometers were in a stable,
resonant configuration.  Additionally, data segments which were flagged as being
of poor quality were not included in the analysis.  For example, data segments 
which were known to have a high rate of seismic transients were excluded from 
the analysis.  After all the data quality cuts were made, there were 28 GRBs left 
to be analyzed for the S2 run, 7 GRBs for S3, and 4 GRBs for S4, for a 
total of 39 GRB triggers.  Of these, 22 GRBs had positions well-localized to 
within a few arcminutes, while 17 GRBs did not.  These 17 GRBs were detected by 
either HETE or IPN.  In the case of HETE, no position measurements were available 
while, in the case of IPN, the GRBs were not well-localized.  Of the 39 GRBs, six
had redshift measurements, four of which were at $z > 1$, and two fell in the 
short-duration category of bursts, i.e. had durations $\leq2$~seconds.  For this 
analysis, due to the small size of the sample, we did not attempt to differentiate 
the GRBs according to their observed properties.  The use of a classification 
scheme in a search can be done in the future with a larger GRB sample.

Information about most of the GRBs was collected from the corresponding GCN
circulars.  The parameters that are relevant for this analysis are the GRB
date and trigger time, and the right ascension and declination.  For those
HETE GRBs which did not have positions, information about the GRB trigger time
was obtained from the HETE website.\footnote{http://space.mit.edu/HETE}  A list
of the GRBs analyzed and relevant information are given in Table 
\ref{tab:GRBsample}.

\begin{table*}
\newcolumntype{f}[1]{D{.}{.}{#1}}
\caption{\label{tab:GRBsample}The GRB sample analyzed}
\begin{ruledtabular}
\begin{tabular}{clf{2}f{2}cf{3}f{3}f{3}f{3}f{7}l}
\multicolumn{1}{c}{LIGO} & \multicolumn{1}{c}{GRB\footnotemark[1]}  & \multicolumn{1}{c}{UTC\footnotemark[2]} & \multicolumn{1}{c}{GPS\footnotemark[3]} & 
            \multicolumn{1}{c}{duration\footnotemark[4]} & \multicolumn{1}{c}{R.A.\footnotemark[5]} & \multicolumn{1}{c}{Dec\footnotemark[6]} & 
                        \multicolumn{1}{c}{$F_{\rm ave}$\footnotemark[7]} & \multicolumn{1}{c}{$F_{\rm ave}$\footnotemark[7]} & 
                                    \multicolumn{1}{c}{time delay\footnotemark[8]} & \multicolumn{1}{c}{IFO\footnotemark[9]} \\
\multicolumn{1}{c}{run} & \multicolumn{1}{c}{date} & \multicolumn{1}{c}{time}               & \multicolumn{1}{c}{time}                & 
            \multicolumn{1}{c}{(seconds)} & \multicolumn{1}{c}{(degrees)}            & \multicolumn{1}{c}{(degrees)}           & 
                        \multicolumn{1}{c}{LHO}                       & \multicolumn{1}{c}{LLO}                       & 
                                    \multicolumn{1}{c}{(seconds)}                  \\\hline
        &             &              &       &         &         &       &       &            \\
S2 & 030215  & 17:11:52    & 729364325.00 &  40   &   ...   &   ...   & ...   & ...   &  ...       & H1,H2 \\
   & 030215a & 11:13:32    & 729342825.00 &  30   &   ...   &   ...   & ...   & ...   &  ...       & H1,H2 \\
   & 030215b & 11:16:28    & 729343001.00 &  40   &   ...   &   ...   & ...   & ...   &  ...       & H1,H2 \\
   & 030216  & 16:13:44    & 729447237.00 &   3   &   ...   &   ...   & ...   & ...   &  ...       & H1,H2 \\
   & 030217  & 02:45:42    & 729485155.00 &  50   & 186.596 & -11.850 & 0.379 & 0.204 &  0.0078867 & H2,L1 \\
   & 030218  & 11:42:38    & 729603771.00 & 200   &   ...   &   ...   & ...   & ...   &  ...       & H1,H2 \\
   & 030221  & 07:46:14    & 729848787.00 & ...   &   ...   &   ...   & ...   & ...   &  ...       & H1,H2 \\
   & 030223  & 09:45:06    & 730028719.00 &  10   &   ...   &   ...   & ...   & ...   &  ...       & H1,H2 \\
   & 030226\footnotemark[10]  & 03:46:31.99 & 730266404.99 &  22   & 173.254 &  25.900 & 0.356 & 0.524 &  0.0059892 & H1,H2,L1 \\
   & 030228  & 20:26:46    & 730499219.00 &  15   &   ...   &   ...   & ...   & ...   &  ...       & H1,H2 \\
   & 030301  & 20:27:20    & 730585653.00 &  30   &   ...   &   ...   & ...   & ...   &  ...       & H1,H2 \\
   & 030308  & 14:06:09    & 731167582.00 & ...   &   ...   &   ...   & ...   & ...   &  ...       & H1,H2 \\
   & 030320a & 10:11:40    & 732190313.00 &  80   & 267.929 & -25.317 & 0.317 & 0.418 &  0.0093172 & H1,H2,L1 \\
   & 030320b & 18:49:17    & 732221370.00 & 150   &   ...   &   ...   & ...   & ...   &  ...       & H1,H2 \\
   & 030323a & 08:42:24    & 732444157.00 &   5   & 297.250 & -12.500 & 0.269 & 0.131 &  0.0088762 & H1,H2,L1 \\
   & 030323b\footnotemark[11] & 21:56:57.60 & 732491830.60 &  25   & 166.525 & -21.900 & 0.533 & 0.336 &  0.0064593 & H1,H2,L1 \\
   & 030324  & 03:12:42.80 & 732510775.80 &  45   & 204.296 &  -0.317 & 0.148 & 0.288 &  0.0086716 & H1,H2 \\
   & 030325  & 14:15:10    & 732636923.00 &   2   &  70.808 & -19.133 & 0.592 & 0.480 &  0.0039660 & H1,H2,L1 \\
   & 030326  & 10:43:41    & 732710634.00 &  10   & 292.967 & -11.717 & 0.191 & 0.407 &  0.0094257 & H1,H2,L1 \\
   & 030329  & 03:31:43    & 732943916.00 & ...   &   ...   &   ...   & ...   & ...   &  ...       & H1,H2 \\
   & 030329a\footnotemark[12] & 11:37:14.67 & 732973047.67 &  22.8 & 161.208 &  21.517 & 0.265 & 0.051 & -0.0095090 & H1,H2 \\
   & 030329b & 15:34:15.35 & 732987268.35 &  65   & 160.626 & -48.572 & 0.635 & 0.665 & -0.0009927 & H1,H2 \\
   & 030331  & 05:38:40.82 & 733124333.82 &  10   & 349.261 &  36.260 & 0.252 & 0.312 & -0.0057539 & H1,L1 \\
   & 030405  & 02:17:28    & 733544261.00 &   5   & 248.275 & -24.150 & 0.565 & 0.377 &  0.0059975 & H1,H2,L1 \\
   & 030406  & 22:42:07    & 733704140.00 &  65   & 285.429 & -68.083 & 0.598 & 0.551 &  0.0014338 & H1,L1 \\
   & 030410  & 11:23:42    & 734009035.00 &   0.3 &   ...   &   ...   & ...   & ...   &  ...       & H1,H2 \\
   & 030413  & 07:34:37    & 734254490.00 &  15   & 198.604 &  62.350 & 0.680 & 0.586 & -0.0031858 & H2,L1 \\
   & 030414  & 13:48:27    & 734363320.00 &  40   & 119.887 & -48.583 & 0.702 & 0.653 &  0.0015308 & H1,H2 \\
\\
S3 & 031108  & 14:11:01    & 752335874.00 &  22   &  66.729 &  -5.930 & 0.278 & 0.313 & -0.0075264 & H1,H2 \\
   & 031109a & 11:11:48    & 752411521.00 &  59   & 327.765 &  20.203 & 0.336 & 0.464 & -0.0088324 & H1,H2 \\
   & 031123  & 22:41:14    & 753662487.00 & ...   &   ...   &   ...   & ...   & ...   &  ...       & H1,H2 \\
   & 031127a & 18:58:58    & 753994751.00 &  10   &   ...   &   ...   & ...   & ...   &  ...       & H1,H2 \\
   & 031127b & 18:59:16    & 753994769.00 &  70   &   ...   &   ...   & ...   & ...   &  ...       & H1,H2 \\
   & 031130  & 02:04:48    & 754193101.00 &   4   &   ...   &   ...   & ...   & ...   &  ...       & H1,H2 \\
   & 031220  & 03:29:56.74 & 755926209.74 &  16.9 &  69.893 &   7.374 & 0.414 & 0.617 &  0.0068643 & H1,H2 \\
\\
S4 & 050223\footnotemark[13]  & 03:09:06    & 793163359.00 &  23   & 271.390 & -62.481 & 0.676 & 0.596 &  0.0027031 & H1,H2 \\
   & 050306  & 03:33:12    & 794115205.00 & 160   & 282.337 &  -9.162 & 0.565 & 0.610 & -0.0013425 & H1,H2,L1 \\
   & 050318\footnotemark[14]  & 15:44:37    & 795195890.00 &  32   &  49.651 & -46.392 & 0.528 & 0.293 &  0.0083075 & H1,H2,L1 \\
   & 050319\footnotemark[15]  & 09:31:18.44 & 795259891.44 &  10   & 154.202 &  43.546 & 0.597 & 0.370 & -0.0070546 & H1,H2,L1 \\
\end{tabular}
\end{ruledtabular}
\footnotetext[1]{For GRBs with the same date, letters are appended to the date 
                 \indent to distinguish the GRBs.}
\footnotetext[2]{UTC time of GRB trigger.}
\footnotetext[3]{GPS time of GRB trigger (seconds since 0h 6 Jan 1980 UTC.)}
\footnotetext[4]{Duration of gamma-ray burst.}
\footnotetext[5]{Right Ascension of GRB.}
\footnotetext[6]{Declination of GRB.}
\footnotetext[7]{Polarization-averaged antenna factor for specified IFO site\\
                 \indent (cf. Eq. \ref{eq:fave}).}
\footnotetext[8]{Time-of-flight of GW signal between LHO and LLO.  
                 A positive \indent value means that the signal arrived first at LLO;
                 a negative value \indent means that the signal arrived first at LHO.}
\footnotetext[9]{Interferometers which were analyzed.}
\footnotetext[10]{$z = 1.986$.}
\footnotetext[11]{$z = 3.372$.}
\footnotetext[12]{$z = 0.168$.}
\footnotetext[13]{$z = 0.5915$.}
\footnotetext[14]{$z = 1.44$.}
\footnotetext[15]{$z = 3.24$.}
\end{table*}

\section{Data analysis}

\subsection{\label{sec:onoff}On-source and off-source data segments}

Since GRBs have well-measured detection times, the search for short-duration
GW signals can be limited to time segments --- called {\it on-source} segments
here --- surrounding the GRB trigger times.  Limiting the search to encompass 
only these time segments significantly reduces the number of search trials, 
compared to a search which makes use of data from the entire run.  In case of a 
detection, such a reduction in trials translates to a larger significance for 
the detection compared to that which would result from an untriggered search.

Making use of on-source segments also means that background estimation can 
proceed by using data stretches --- called {\it off-source} segments here ---
which are outside the on-source segments, but which are still close enough
in time to the on-source segments so that the off-source data are similar in
character to, and representative of, the on-source data.

In this analysis, the length of each on-source segment was chosen to be 180
seconds, with the first 120 seconds of the LIGO on-source data occurring before
the GRB trigger time, and the last 60 seconds occurring after the trigger time.
This window length is longer than the expected time delay between a
gravitational-wave signal and the onset of a GRB signal, which is of the order
of several seconds \cite{rees94,kochanek93,piran99}, but which in certain models
can be as large as 100 seconds \cite{meszaros06}.  The large search window 
also takes into account the uncertainty in the definition of the measured GRB 
trigger time, i.e. it takes into account the possibility that the trigger time 
used in the analysis occurred before or after the actual start of a gamma-ray 
burst signal.  Many gamma-ray light curves show sub-threshold, precursor bursts 
which occur before the measured GRB trigger time, hence our choice of an asymmetric 
search window around the trigger time.

For each GRB, a search for a GW signal was carried out using data from each
{\em pair} of IFOs that was operating properly at that time.  Additionally,
LHO-LLO on-source pairs were analyzed only when GRBs had well-defined 
positions, since position information is necessary to calculate the LHO-LLO 
time-of-flight delay.  After all the data quality cuts were made, there were 
59 IFO-IFO on-source pairs that were analyzed.  This number is larger than the 
number of GRB triggers because, for each GRB trigger, it was possible to have up 
to three IFO pairs pass the data quality cuts.  There were 35 H1-H2 on-source 
pairs analyzed, 12 for H1-L1, and 12 for H2-L1.

The software used in this analysis is available in the LIGO Scientific
Collaboration's CVS archives with the tag multigrb\_r1 in 
MATAPPS.\footnote{http://www.lsc-group.phys.uwm.edu/cgi-bin/cvs/viewcvs.cgi
/matapps/src/searches/burst/multigrb\\
/?cvsroot=lscsoft\&sortby=rev\#dirlist}

\subsection{\label{sec:datacond}Data conditioning}

Before the crosscorrelation between two LIGO data streams was calculated, the
time series data from each interferometer was conditioned.  This consisted of
whitening, phase-correction, and bandpassing from 40~Hz to 2000~Hz.  The 
sampling rate was retained at 16384 samples/s.  Whitening was done to make
sure the resulting spectrum of the data was flat instead of being dominated 
by low-frequency or high-frequency components.  The procedure consisted of 
using one-second data units to whiten the adjacent one-second data and, as 
a consequence, removed any non-stationarity in the data having a time scale 
larger than one second.  The whitening procedure also removed known lines.

The response functions of the three LIGO interferometers to a given GW strain
signal are not exactly the same.  A GW signal impinging on the three 
interferometers will thus appear as having slightly different phases in the 
corresponding time series data (even after correcting for the LHO-LLO time-of-flight 
delay).  Phase correction of the time series data was therefore done to remove the 
differences that can be attributed to the different response functions of the 
interferometers.  The phase correction process made use of the measured, 
time-dependent, response functions of the interferometers.

\subsection{\label{sec:ccstat}Measuring the crosscorrelation statistic}

The search method consisted of a simple ``binned'' search in which the 
180-second conditioned on-source time-series for each IFO was divided into time 
intervals (or bins) and the crosscorrelation for each IFO-IFO time bin pair 
calculated.  Crosscorrelation bins of lengths 25~ms and 100~ms were used to 
target short-duration GW signals with durations of $\sim$~1~ms to $\sim$~100~ms.
These crosscorrelation lengths were found, through simulations, to provide
sufficient coverage of the targeted short-duration GW signals.  Using bins
much shorter than 25 ms would considerably increase the trials in the search,
and therefore decrease the significance of a candidate GW event, while using
bins much longer than 100 ms would considerably diminish the crosscorrelation
strength of signals in the two data streams due to the increased duration of
noise.  The crosscorrelation, $cc$, is defined as:
\begin{equation}
cc = \frac{\displaystyle \sum_{i=1}^{m} [s_1(i)-\mu_1][s_2(i)-\mu_2]} {\sqrt[]
          {\displaystyle \sum_{j=1}^{m} [s_1(j)-\mu_1]^2}\; \sqrt[]
          {\displaystyle \sum_{k=1}^{m} [s_2(k)-\mu_2]^2}}
\label{eq:ccdef}
\end{equation}
where $s_1$ and $s_2$ are the two time series to be correlated, $\mu_1$ and 
$\mu_2$ are the corresponding means, and $m$ is the number of samples in the
crosscorrelation, i.e. the crosscorrelation integration length multiplied by the
sampling rate of 16384~samples/s.  The possible values of the normalized
crosscorrelation range from $-1$ to $+1$.

The bins were overlapped by half a bin width to avoid inefficiency in detecting
signals occurring near a bin boundary.  The crosscorrelation value was calculated
for each IFO-IFO bin pair and, for each crosscorrelation bin length used, the 
largest crosscorrelation value --- in the case of an H1-H2 search --- obtained 
within the 180-second search window was considered the most significant 
measurement for that search, for that crosscorrelation bin length, for that IFO 
pair.  In the case of an H1-L1 or H2-L1 search, it was the largest {\it absolute} 
value of the crosscorrelations that was taken as the most significant 
measurement.  This was done to take into account the possibility that signals at 
LHO and LLO could be anti-correlated depending on the gravitational wave's 
(unknown) polarization.  In the sections that follow, a reference to the ``largest
crosscorrelation'', in the case of an LHO-LLO analysis, will always mean the 
largest absolute value of crosscorrelations.

For those GRBs which had well-defined positions, the position of the GRB in the
sky at the time of the burst was used to calculate the GW signal's time-of-flight
delay between the LHO and LLO observatories.  Each LHO-LLO pair of 180-second
on-source segments were shifted in time relative to each other by the
corresponding time-of-flight amount before the crosscorrelations were calculated.
For those GRBs which were not well-localized, only H1-H2 on-source pairs were
analyzed.  For these GRBs, the maximum uncertainty in the LHO-LLO time delay is 
$\pm 10$ ms, which is of the same scale as the signal durations targeted by the 
analysis, and such a time offset between signals at the two interferometers would 
have a considerable effect on the measured crosscorrelation.

\subsection{Post-trials distributions}

To estimate the significance of the loudest event, i.e. the largest 
crosscorrelation, that was found in an on-source segment corresponding to a GRB 
and an IFO pair, we used off-source data within a few hours of the on-source 
data to measure the crosscorrelation distribution of the noise.  This 
distribution was obtained for each GRB, for each IFO pair, for each 
crosscorrelation length by applying the search (described in 
Sections~\ref{sec:datacond} to \ref{sec:ccstat}) on the off-source segments.
The total length of the off-source region was about three hours surrounding the
on-source segment.  Each distribution was constructed by collecting the largest
crosscorrelation (or largest absolute value of crosscorrelations, in the case of
H1-L1 and H2-L1) from each 180-second segment of the off-source region.  This
{\it post-trials} distribution takes into account the number of effective trials
that was used in searching the on-source segment.

To obtain enough statistics for each distribution, time shifts were performed
such that the time series of each IFO was shifted by multiples of 180 seconds
relative to the other IFO and two 180-second stretches from the two IFOs were
paired at each shift, making sure that two 180-second time stretches were paired
only once for each distribution.  The time shift procedure effectively increased
the length of the off-source data to about 50 hours or more, typically.

\begin{figure}
\includegraphics[width=3.40in]{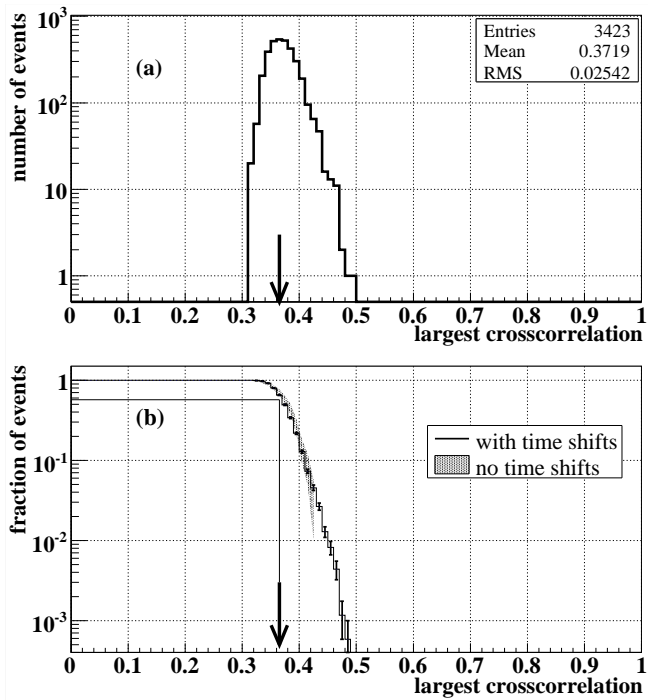}
\caption{\label{fig:posttrials}(a) Example of a crosscorrelation post-trials
distribution for the 25-ms crosscorrelation window, for the H1-H2 IFO pair.
Data from off-source segments for GRB 050318 was used.  (b) Cumulative
distribution of (a), normalized to the total number of entries in the
distribution.  Both distributions with and without time shifts are shown,
including the statistical errors.  The arrow points to the largest 
crosscorrelation found in the on-source segment for GRB 050318.  In this
example, the largest crosscorrelation of 0.36 has a local probability of 0.57.}
\end{figure}

As an example, the post-trials distribution for GRB 050318, for the H1-H2 IFO
pair and for the 25-ms crosscorrelation length, is shown in
Fig.~\ref{fig:posttrials}.  For comparison, the cumulative plot shows both the
distribution obtained with time shifts, and the distribution obtained without 
employing time shifts.

Each resulting post-trials distribution was used to estimate the cumulative
probability that the largest crosscorrelation found in the corresponding
on-source segment could be due to noise.  This was done by determining what
fraction of the distribution were at least as large as the loudest
crosscorrelation found in the on-source segment.  For example, the significance
of the loudest 25-ms crosscorrelation found in the H1-H2 on-source segment of
GRB 050318, indicated by an arrow in Fig.~\ref{fig:posttrials}(b), can be
estimated by using the plotted post-trials distribution.  This probability will
be referred to interchangeably in this paper as the post-trials, or {\it local},
probability of the on-source crosscorrelation statistic.  This is also known in
the literature as the {\it false alarm} probability.

Since H1 and H2 are colocated, environmental disturbances can give rise to
correlated transient noise in the two interferometers.  The effect of these
correlated environmental noise on an H1-H2 crosscorrelation were, however,
suppressed by:  the judicious use of data quality cuts (cf. Section III), the
applied data conditioning (cf. Section IVB), and the use of off-source data
immediately surrounding the on-source data to estimate the background noise
(cf. this section), which made it more likely that the background would
properly reflect the rate of any correlated noise in the on-source data.

\begin{figure}[t]
\includegraphics[width=3.40in]{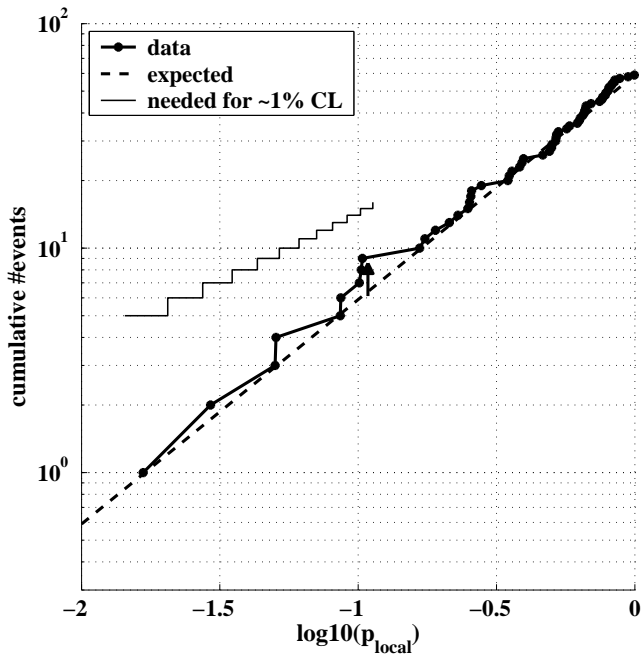}
\caption{\label{fig:localp25ms}Cumulative local probability distribution
resulting from the search of 59 IFO-IFO on-source pairs using a 25-ms
crosscorrelation length.  The most significant excess is indicated by the
arrow.  The expected distribution under the null hypothesis is indicated
by the bold, dashed line. The excess needed for a $\sim1\%$ confidence in
the null hypothesis is indicated by the solid line.  The maximum excess
indicated by this line is 15 events because only the 15 most significant
events in the actual distribution are tested.}
\end{figure}

\begin{figure}[t]
\includegraphics[width=3.40in]{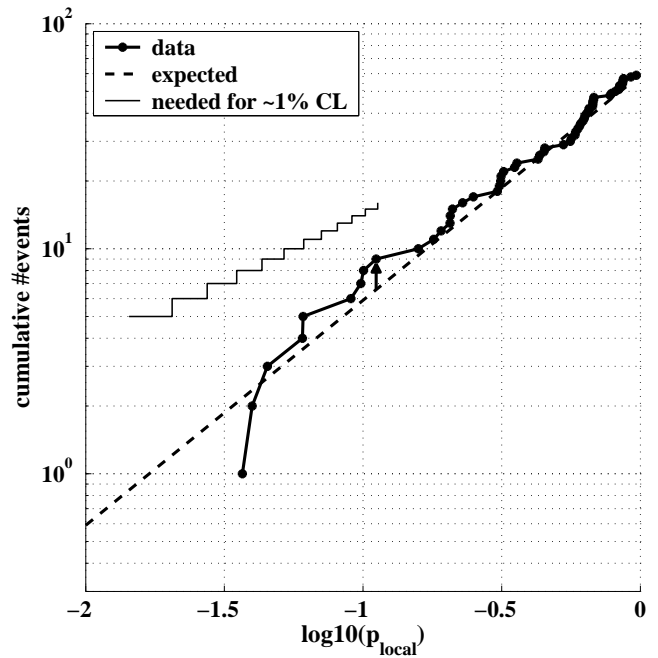}
\caption{\label{fig:localp100ms}Similar to Fig. \ref{fig:localp25ms}, but
using a 100-ms crosscorrelation length.}
\end{figure}

The cumulative distribution of local probabilites resulting from the search of
59 on-source segment pairs is shown in Fig.~\ref{fig:localp25ms} for the 25-ms
crosscorrelation length, and in Fig.~\ref{fig:localp100ms} for the 100-ms
crosscorrelation length.  Also shown (bold dashed lines) is the expected 
distribution under a null hypothesis.  There were no loud events that were not 
consistent with the expected distribution, and we therefore conclude that there
was no loud GW signal associated with any single GRB in the sample.

\section{\label{sec:stattest}Statistical tests}

As mentioned earlier, GW signals from individual GRBs are likely to be weak
in most cases due to the cosmological distances involved.  Therefore, besides
searching for GW signals from each GRB, we also consider the detection of a GW
signature associated with a sample of several GRBs.  Such approaches, first
proposed in the context of GWs in \cite{FMR}, have already been used
\cite{astone1,astone2} to analyze resonant mass detector data using triggers
from the BATSE and BeppoSAX missions.

We use two different statistical methods to look for a GW signature associated 
with a sample of multiple GRBs.  As one may expect, the statistical performance 
of a method will depend on the nature of the underlying source population 
distribution.  The two different methods presented here have complementary 
properties in this respect.  The first statistical test presented, the binomial 
test, is most effective when several events contribute to the tail, i.e. the 
significant end, of the probability distribution of a sample.  Moreover, it is 
also effective when there is a single significant event in the sample.  The 
second test, the rank-sum test, is more effective at detecting the cumulative 
effect of weaker signals, but it is not very effective at detecting a few large 
events which fall on the tail of a probability distribution.

Since the signal strengths targeted by these two methods are slightly different,
the resulting significances from the two methods can be different when there 
are real signals present in the sample.  If a detection is claimed and the more 
significant measurement from the two statistical tests is chosen, then the 
proper statistical treatment, in order to arrive at a final significance, would 
be to impose a penalty factor for using two statistical tests to search for the
cumulative signal.

\subsection{Testing a probability distribution:  The binomial test}

Under a null hypothesis, the distribution of local probabilities is expected to
be uniformly distributed from 0 to 1.  The measured distribution of local
probabilities was tested to search for an excess which may have been due to the
cumulative effect of weak GW signals.  In particular, we searched the tail of
the distribution, or the smallest probabilities found in the on-source searches,
by using the {\it binomial test}.  To test the tail of a probability distribution,
one first makes a choice as to how many events, $n$, in the tail would be tested
out of the total number of events, $N$, in the sample.  In this analysis, there
were 59 IFO-IFO on-source pairs, and the upper 25\% of the resulting probability
sample, or the 15 most significant events, was tested.  The probabilities of
these $n$ events are then sorted according to increasing value, i.e.  decreasing
significance:  $p_1, p_2, p_3, ..., p_i, ..., p_n$.  For each of these
probabilities, $p_i$, one calculates the cumulative binomial probability, which
is the probability for getting $i$ or more events at least as significant as
$p_i$:
\begin{align}
P_{\ge i}(p_i) & = P_i(p_i) + P_{i+1}(p_i) + P_{i+2}(p_i) + ... + P_{N}(p_i) \\
               & = 1 - [P_0(p_i) + P_1(p_i) + P_2(p_i) + ... + P_{i-1}(p_i)]
\label{eq:pbinocum}
\end{align}
and where $P_i(p)$ is the binomial probability for getting $i$ successes in $N$
trials:
\begin{align}
P_{i}(p) & = \frac{N!}{i!(N-i)!} p^i (1-p)^{N-i}
\label{eq:pbino}
\end{align}
Here, $N$ is the number of on-source searches, which is 59, and ``success'' means
getting $i$ events at least as significant as $p$.  Note that if there is one
loud event in the sample, with $p \ll 1$, then it follows from
Eqs.~\ref{eq:pbinocum} and \ref{eq:pbino} that the cumulative binomial
probability is,
\begin{align}
P_{\ge 1}(p) & = 1 - (1-p)^N \\
             & \approx Np
\end{align}
Thus, the binomial test is able to automatically handle the case of a single
loud event in the distribution.

After the cumulative binomial probability, $P_{\ge i}(p_i)$, has been calculated
for each post-trials probability, $p_i$, the smallest binomial probability in the
set is identified.  This smallest binomial probability will point to the most
significant excess that was found in searching the tail of the probability
distribution.

The most significant excess that was found by the binomial test in the tail of
the distribution is indicated by an arrow in Figs. \ref{fig:localp25ms} and 
\ref{fig:localp100ms}.  For the 25-ms distribution, the smallest binomial 
probability found was $P_{\ge  9}(p_9 = 0.104) = 0.153$.  This means that the 
binomial test found that the most significant excess in the tail of the 
distribution consisted of nine events with local probabilities $p \le 0.104$, 
and that the binomial probability for having nine or more events at least as 
significant as $0.104$, given 59 trials, is $0.153$.

In the case of the 100-ms distribution, the smallest binomial probability found
was $P_{\ge  9}(p_9 = 0.112) = 0.207$.  This means that the binomial test found
that the most significant excess in the tail of the distribution consisted of
nine events with local probabilities $p \le 0.112$, and that the binomial
probability for having nine or more events at least as significant as $0.112$,
given 59 trials, is $0.207$.

Searching the tail of a post-trials probability distribution for the most
significant excess introduces additional trials to the search.  We thus need
to test the most significant excess found in the tail of each local probability 
distribution against the null hypothesis to properly establish its level of 
significance.  The expected distribution of the binomial probability statistic 
under the null hypothesis was obtained through simulations.  The simulations 
consisted of randomly generating 59 numbers uniformly distributed from 0 to 1 to 
simulate 59 post-trials probabilities under the null hypothesis.  Then the same 
binomial test that was applied to the actual post-trials probability distribution 
was applied to this distribution of random events to search for the most 
significant excess in the 15 most significant events in the tail.  This was 
repeated a million times, and the binomial probability of the most significant 
excess found in each trial was collected.  The resulting distribution of binomial 
probabilities under the null hypothesis, in effect, takes into account the 
number of trials used in searching the tail of the post-trials distribution.

Results of these simulations show that, under the null hypothesis, the 
probability for getting a measurement at least as significant as 0.153 that was 
found in the 25-ms search is 0.48.  In other words, under the null hypothesis, 
1 in 2.1 sets of 59 on-source searches will result in a most significant excess 
with a binomial probability at least as significant as 0.153.  This quantifies 
the conclusion that the result of the 25-ms search is consistent with the null 
hypothesis.

Similarly, we find that, under the null hypothesis, the probability for getting
a measurement at least as significant as 0.207 that was found in the 100-ms
search is 0.58.  In other words, under the null hypothesis, 1 in 1.7 sets of
59 on-source searches will result in a most significant excess with a binomial
probability at least as significant as 0.207.  And, as with the 25-ms result,
this level of significance for the 100-ms search result is consistent with the
null hypothesis.

Also shown in Figs. \ref{fig:localp25ms} and \ref{fig:localp100ms} is a curve
indicating the excess needed for a $\sim1\%$ confidence in the null hypothesis.  
At each local probability, the curve gives the cumulative number of events 
needed to obtain a $\sim1\%$ final probability under the null hypothesis, given 
59 on-source pairs.

\subsection{Maximum likelihood ratio based tests}

A maximum likelihood ratio test~\cite{helstrom} for detecting a GW signature
associated with a sample of multiple triggers was derived in~\cite{SDM:gwdaw9}.
(It was shown there that~\cite{FMR} is a special case of the maximum likelihood 
ratio approach.)  The method proposed in~\cite{SDM:gwdaw9} cannot be applied 
directly to the entire GRB sample described above since the largest 
crosscorrelation values were obtained in different ways for H1-H2 and H1-L1 
(H2-L1) (cf. Section~\ref{sec:ccstat}).  In the following, we will only use the 
largest crosscorrelations from H1-H2 on-source segments.  This reduces the total 
number of GRB on-source segments used in this test to 35.

Let the largest crosscorrelation from the $i^{\rm th}$ GRB on-source segment be 
denoted as $cc_{\rm max,i}$.  If we do not use any prior probability distribution for 
the properties of GW signals associated with GRBs, the maximum likelihood ratio 
detection statistic is simply the average of the largest crosscorrelation values 
from the GRB set,
\begin{equation}
\chi =\frac{1}{N_{\rm GRB}} \sum_{i} cc_{{\rm max},i} ~~~,
\label{lrtest}
\end{equation}
where $N_{\rm GRB}$ is the number of H1-H2 GRB on-source segments used.  We call 
$\chi$ the {\em sum-max} statistic.

To build in robustness against instrumental noise artefacts, such as short 
duration transients, we replace the sum-max statistic, which was derived for the 
ideal case of Gaussian and stationary noise, by a non-parametric counterpart. 
The on-source and off-source largest crosscorrelation values are pooled into two 
separate sets and the Wilcoxon rank-sum test \cite{ranksum} is used for the null 
hypothesis that the two sets of samples were drawn from the same underlying true 
distribution.

\begin{figure}
\includegraphics[width=3.75in]{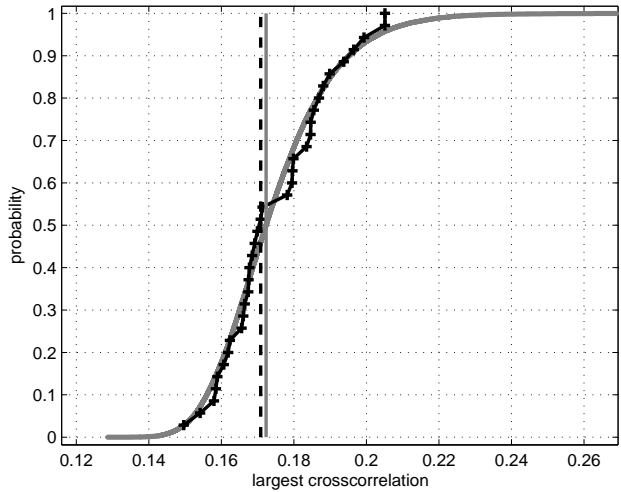}
\caption{The cumulative distributions of the on-source (solid black with 
$+$ marker) and off-source (solid gray) largest H1-H2 crosscorrelations from the 
100-ms search.  The vertical lines denote the locations of the medians of the 
off-source (gray) and on-source (black, dashed) samples.
\label{ranksum_sig_fig}}
\end{figure}

\begin{figure}
\includegraphics[width=3.75in]{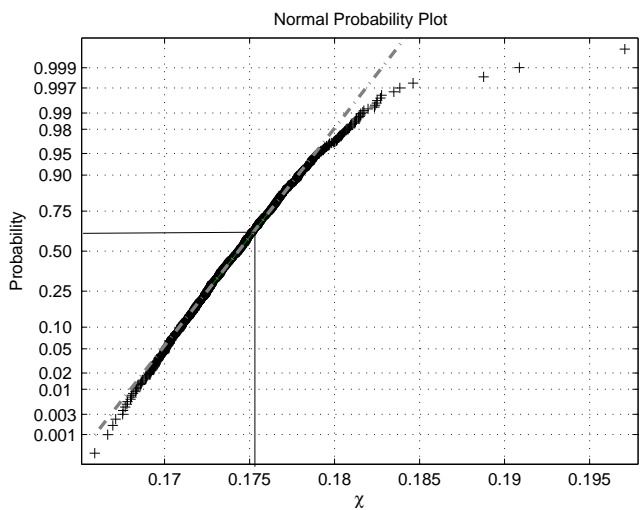}
\caption{Plot of the cumulative distribution function of the off-source values
of the sum-max statistic $\chi$.  The dashed line shows the best fit normal 
distribution.  The solid horizontal and vertical lines indicate the location of 
the on-source values of $\chi_{\rm on}$ and the corresponding cumulative 
probability.  The on-source value was $\chi_{\rm on} = 0.1753$, which yields a 
cumulative probability of 0.64 or an empirical significance of 1 - 0.64 = 0.36.
\label{empirical_sig_fig}}
\end{figure}

The cumulative distribution of the on-source and off-source largest 
crosscorrelations from the 100-ms search are shown in 
Fig.~\ref{ranksum_sig_fig}.  Application of the rank-sum test shows that the 
significance of the null hypothesis is $0.64$.  This implies that one out of 
$1.6$ trials can show a false positive detection at this significance threshold.
Assuming that GRB triggers occur at a rate of one per day, one year of observation
would contain approximately 10 collections of 35 GRBs.  In order to achieve a 
low false detection probability, we would require a much lower significance, 
such as $\leq 0.01$, in order to reject the null hypothesis.

As a further check, we also compute the empirical significance of the on-source
value of $\chi$ with respect to the set of off-source $\chi$ values.  Values of 
the off-source $\chi$ statistic were calculated by pooling the largest 
crosscorrelations from the H1-H2 off-source segments, then dividing this pool 
into subsets, each of which had $N_{\rm GRB}$ number of elements.  For each of 
these subsets, the $\chi$ statistic was calculated using Eq. \ref{lrtest}.  The 
empirical significance is defined as the fraction of off-source $\chi$ values 
greater than or equal to the on-source $\chi$ value.  The empirical significance 
has more scatter than a significance computed with a {\em known} null distribution  
since we only have a finite number of off-source values.  However, the number of 
off-source values in this analysis is large enough that we can ignore the associated 
error.

The mean and standard deviation of the off-source sum-max sample were $0.1744$
and $0.0029$, repectively.  The on-source  value of sum-max was 
$\chi_{\rm on}=0.1753$.  Figure~\ref{empirical_sig_fig} shows the distribution of 
the off-source values of the test statistic.  The empirical significance of the 
null hypothesis is $0.36$.  Following the discussion above, it is clear that 
this result is consistent with no detection.

\section{\label{sec:grblimits}Single-GRB limits}

Simulations were done to estimate the sensitivity of the search method to 
incident GW burst signals.  This process was limited by the fact that the
theoretical waveforms of the GW burst signals associated with GRBs were not
known.  Other unknown quantities were: the polarization of the waves, the
orientation of the source relative to the observer, and the redshifts of most of
the GRBs.  Conscious of these limitations, we proceed to set upper limits on the 
root-sum-square amplitude ($h_{\rm rss}$) of GW burst signals incident on the
interferometers during the on-source times by using simulated waveforms with
burst-like characteristics, adding these waveforms to the raw IFO data streams,
and measuring the resulting crosscorrelations.

The antenna response of an IFO to incident, independent gravitational-wave 
strains, $h_+(t)$ and $h_\times(t)$, depends on the relative position of the 
source in the sky and the polarization of the wave \cite{300years}:  
\begin{equation}
h(t) = F_+(\theta,\phi,\psi)h_+(t) + F_{\times}(\theta,\phi,\psi)h_\times(t) \,,
\label{eq:hresponse}
\end{equation}
where $(\theta,\phi)$ is the position of the source relative to the IFO's zenith 
and x-arm, respectively; $\psi$ is the polarization angle of the gravitational-wave; 
and $F_+(\theta,\phi,\psi)$ $F_\times(\theta,\phi,\psi)$ are the corresponding 
``plus'' and ``cross'' antenna factors.  For most of the GRBs analyzed, the position, 
$(\theta,\phi)$, was known.  The polarization angle, $\psi$, however, was an unknown
parameter for all of the GRBs.  Since the antenna factor is used in the simulations, 
upper limits were not set for GRBs which did not have well-defined positions.  The
polarization-averaged antenna factor is defined as:
\begin{equation}
F_{ave}(\theta,\phi) = \sqrt{\frac{F_+^2 + F_{\times}^2}{2}} = 
                       \sqrt{\bigl<F_+^2\bigr>_\psi} = \sqrt{\bigl<F_\times^2\bigr>_\psi} \,.
\label{eq:fave}
\end{equation}

We used sine-gaussians as the simulated waveforms for $h_+(t)$ and cosine-gaussians
for $h_\times(t)$ in Eq.~\ref{eq:hresponse}:
\begin{eqnarray}
h_+(t) &=\, h_{+,0} \sin(2 \pi f_0 t) \exp\biggl(\dfrac{-(2\pi f_0 t)^2}{2Q^2}\biggr) \,, \\
h_{\times}(t) &=\, h_{\times,0} \cos(2 \pi f_0 t) \exp\biggl(\dfrac{-(2\pi f_0 t)^2}{2Q^2}\biggr) \,,
\label{eq:sg}
\end{eqnarray}
where $f_0$ is the central frequency of the sine-gaussian and cosine-gaussian,
$h_{+,0}$ and $h_{\times,0}$ are the amplitude parameters of the $+$ and $\times$
polarization signals, respectively, and $Q$ is a dimensionless constant which
represents roughly the number of cycles with which the waveform oscillates with
more than half of the peak amplitude.  The root-sum-squared (rss) amplitude of 
$h_+(t)$ and $h_{\times}(t)$ is related to these parameters via:
\begin{eqnarray}
\label{eq:rssdef1} \sqrt{\int |h_+(t)|^2 ~ dt} &\approx\, 
                   h_{+,0}\, \sqrt{\dfrac{Q}{4 \sqrt{\pi} f_0}} \quad \text{for} \quad Q \gtrsim 3 \,, \\
\label{eq:rssdef2} \sqrt{\int |h_{\times}(t)|^2 ~ dt} &\approx\,  
                   h_{\times,0}\, \sqrt{\dfrac{Q}{4 \sqrt{\pi} f_0}} \quad \text{for} \quad Q \gtrsim 3 \,.
\end{eqnarray}

Using these waveforms for $h_+(t)$ and $h_\times(t)$, we simulated 
circularly polarized GW waves by setting the sine-gaussian and cosine-gaussian 
amplitudes equal to each other, $h_{+,0} = h_{\times,0} \equiv h_0$.  To 
simulate linearly polarized waves, we set $h_{\times,0} = 0$.  In the discussion
that follows, we define the $h_{\rm rss}$ of a simulated waveform as:
\begin{equation}
h_{\rm rss} = \sqrt{\int (|h_+(t)|^2 + |h_\times(t)|^2) ~ dt} \quad.
\label{eq:hrss}
\end{equation}

Since the polarization angle, $\psi$, was not known for any GRB, a random 
polarization angle from 0 to 360 degrees was generated for each simulated 
waveform event.  In the case of LHO-LLO simulations, the source 
position-dependent difference in the polarization angles at LHO and LLO --- due 
to the non-aligned detector arms --- was taken into account.  Finally, before 
adding the attenuated waveform given by Eq.~\ref{eq:hresponse} into an IFO's raw 
data stream, it was first calibrated using the measured response function of the 
IFO.

\begin{figure}[t]
\includegraphics[width=3.40in]{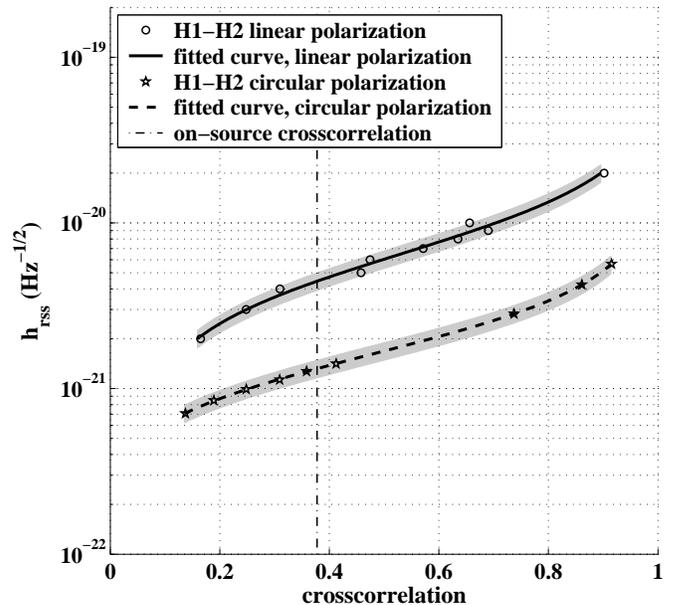}
\caption{\label{fig:ulcurve25ms}Examples of upper limit curves that were used to
set upper limits on $h_{\rm rss}$ using linear and circular polarizations.  These 
were the upper limit curves for the H1-H2 IFO pair, for GRB~050306, using 
sine-gaussians with Q~=~8.9 and $f_0=$~150~Hz.  The shaded regions indicate the
total $1\sigma$ uncertainty in the $h_{\rm rss}$ value.}
\end{figure}

Following the procedure outlined above, simulated sine-gaussians with different
frequencies and $h_{\rm rss}$ values were added to each IFO on-source data at known
times.  Randomness in the injection times of the order of the crosscorrelation
length was introduced to simulate the fact that the relative time of the GW
signal within the 180-second search window was not known.  Data with injected
signals were then conditioned using the procedure outlined in
Section~\ref{sec:datacond}.  The search was then applied to the data near the
injection times --- not to the entire 180-second on-source segment --- to find 
the largest crosscorrelations around the injection times.  This simulation
procedure resulted in the determination of the probability density, 
$p(cc|h_{\rm rss})$, for measuring a crosscorrelation, $cc$, corresponding to a 
signal injected in an on-source segment with a certain $h_{\rm rss}$ value.

The method used to set upper limits on $h_{\rm rss}$ follows the standard recipe 
for setting frequentist upper limits \cite{pdg02}.  If $p(cc|h_{\rm rss})$
is the probability density for measuring a crosscorrelation, $cc$, in an on-source 
segment given a signal with a certain $h_{\rm rss}$ value, then the 90\% upper 
limit curve can be constructed from the set $(cc_{90},h_{\rm rss})$, such that,
\begin{align}
\label{eq:ul1}0.90 & = \int_{cc_{90}}^\infty p(cc|h_{\rm rss}) \; d(cc) ~~~.
\end{align}

Examples of upper limit curves obtained through this procedure are shown in
Fig.~\ref{fig:ulcurve25ms}, with one curve corresponding to linear polarization,
and the other curve corresponding to circular polarization.  These curves were 
obtained using the H1-H2 on-source data for GRB~050306; 150-Hz, Q = 8.9 
sine-gaussians; and a 25-ms crosscorrelation length.  Each curve shows the 
$h_{\rm rss}$ value of the simulated waveform versus $cc_{90}$, the crosscorrelation 
value at which 90\% of the measured crosscorrelation values were larger 
(see Eq.~\ref{eq:ul1}).  The data was fitted with a four-parameter sigmoid 
function, 
\begin{equation}
cc_{\rm 90} = p_1 + \frac{1-p_1}{p_4\bigl(1+\exp[-p_2(\log_{10}({h}_{\rm rss})-p_3)]\bigr)} \quad,
\label{eq:ulfit}
\end{equation}
where parameter $p_1$ defined the asymptote of $cc_{\rm 90}$ at small values of 
${h}_{\rm rss}$, $p_4$ tracked the asymptote of $cc_{\rm 90}$ at large values 
of ${h}_{\rm rss}$ (i.e. $p_4 \approx 1/\text{asymptote}$), $p_3$ was the value 
of ${h}_{\rm rss}$ which gave a mid-range value of $cc_{\rm 90}$, and $p_2$ 
defined the slope of the curve.  The largest crosscorrelation found in the 
on-source segment is also shown in Fig.~\ref{fig:ulcurve25ms} (vertical dashed 
line).  The 90\% $h_{\rm rss}$ upper limit, before uncertainties, was found by 
evaluating the upper limit curve, which is the inverse of Eq.~\ref{eq:ulfit}, at 
the largest on-source crosscorrelation value found in the search.

The curves in Fig.~\ref{fig:ulcurve25ms} also show the estimated total $1\sigma$
uncertainty in the measurement of the $h_{\rm rss}$ values.  The uncertainty in 
the $h_{\rm rss}$ values comes from measured random and sytematic errors in the 
calibration parameters that were used to calibrate the simulated waveforms, and
also from the statistical errors which come from the simulation procedure.
Depending on which science run and IFO pair is being considered, the total
$1\sigma$ uncertainty from all these sources ranged from $\sim10\%$ to 
$\sim13\%$.  However, for GRB~030217 and GRB~030226, the total uncertainty 
was about $\sim22\%$ for the H1-H2 and H1-L1 IFO pairs, due to larger 
calibration errors during the times of those GRBs.  The final 90\% $h_{\rm rss}$ 
upper limits were obtained by adding the corresponding total $1.28\sigma$ 
uncertainties to the values obtained from the upper limit curves.

\begin{figure}
\includegraphics[width=3.45in]{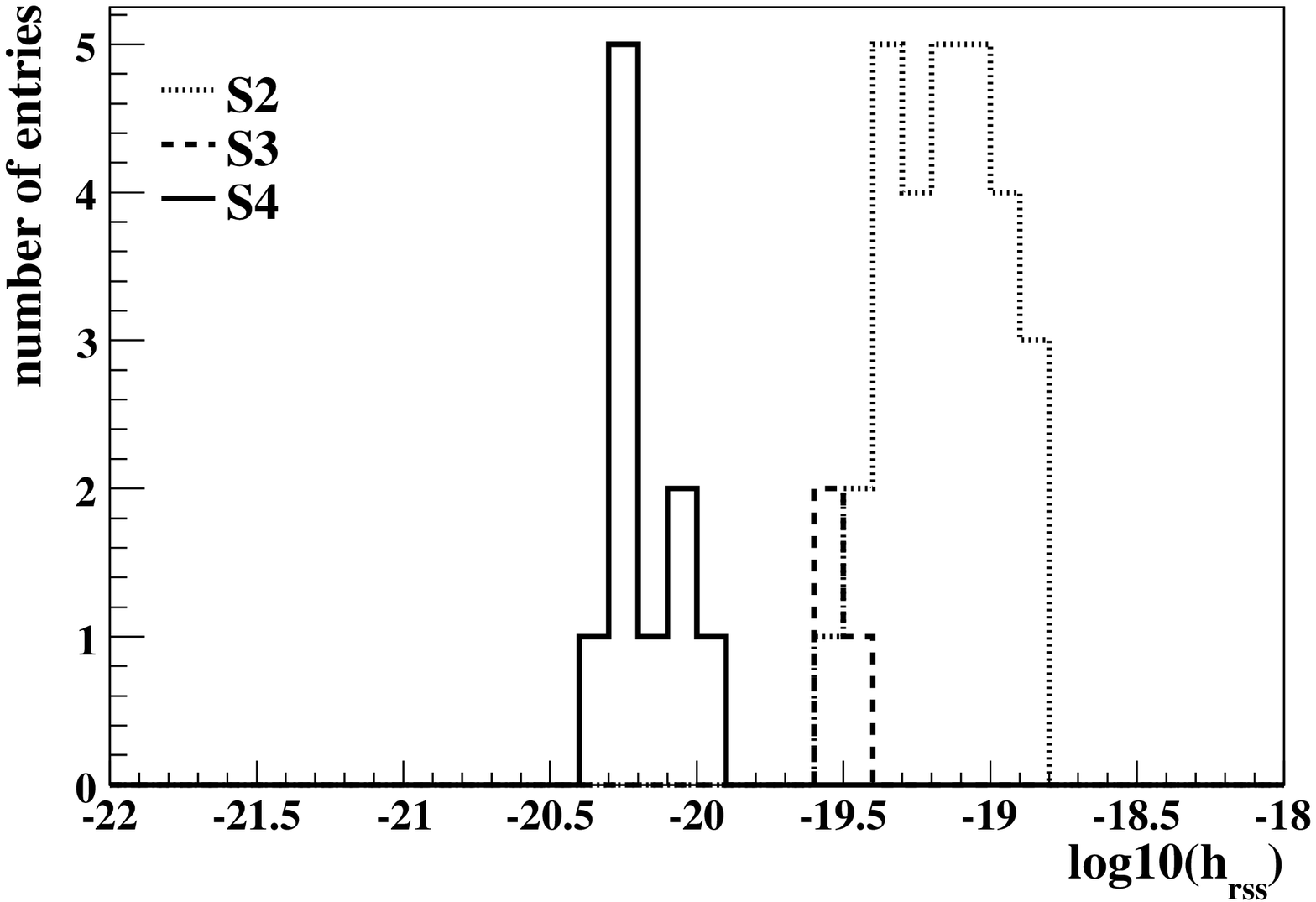}
\caption{\label{fig:ulallpsil}Progression of $h_{\rm rss}$ upper limits from the S2 to
S4 LIGO runs for linearly polarized sine-gaussian waveforms; \mbox{25-ms} crosscorrelation.}
\end{figure}

\begin{figure}
\includegraphics[width=3.45in]{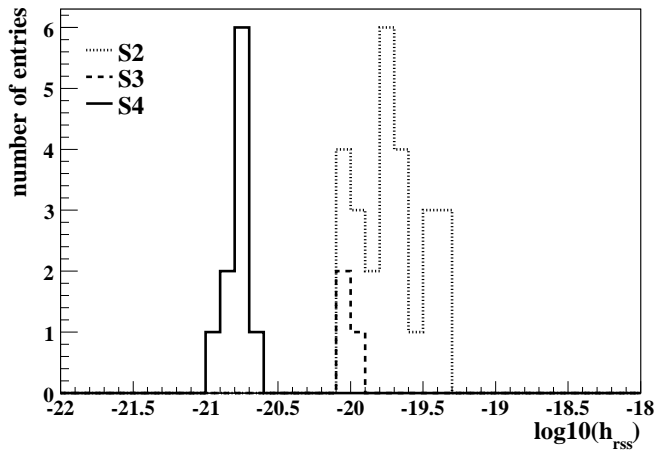}
\caption{\label{fig:ulallpsic}Progression of $h_{\rm rss}$ upper limits from the S2 to
S4 LIGO runs for circularly polarized sine-gaussian waveforms; 25-ms crosscorrelation.}
\end{figure}

The upper limits resulting from the use of Q = 8.9 sine-gaussians and a 25-ms
crosscorrelation length, for GRBs with well-localized positions, are listed in
Tables~\ref{tab:s4ul25ms} to \ref{tab:s2ul25ms} for linearly polarized waveforms,
and in Tables~\ref{tab:s4ul25mspsic} to \ref{tab:s2ul25mspsic} for circularly 
polarized waveforms.  Corresponding limits from the use of a 100-ms 
crosscorrelation length are listed in Tables~\ref{tab:s4ul100ms} to 
\ref{tab:s2ul100ms}, and in Tables~\ref{tab:s4ul100mspsic} to 
\ref{tab:s2ul100mspsic}.  It can be seen that the upper limits for the two 
crosscorrelation lengths do not differ much for the waveforms that were used.  
The upper limits for $f_0 = 250$ Hz and 25-ms crosscorrelation length are 
plotted in Figs.~\ref{fig:ulallpsil} and \ref{fig:ulallpsic} for linear and 
circular polarizations, respectively.  The improvement in sensitivity from the 
S2 to S4 runs can be seen in these plots.  The best upper limits from the three 
science runs are given in Table~\ref{tab:ulbest}.  From the S2 to the S4 run, 
there was an improvement in sensitivity by about an order of magnitude.

It can also be seen from Figs.~\ref{fig:ulallpsil} and \ref{fig:ulallpsic}
that, for most of the GRB source positions, the circular polarization limits are
better than the linear polarization limits by about a factor of 3.5.  This is
always true in the case of H1-H2 upper limits since waveforms at the two 
co-aligned LHO IFOs were always in phase (after calibrations).  For LHO-LLO 
upper limits, there were two cases, GRB 030217 and 030323a, in which the positions 
of the GRBs relative to the IFOs were such that circularly polarized waveforms at 
LHO and LLO were sufficiently out of phase so that upper limits for circular 
polarization were not determinable for those GRBs.

\begin{table}
\caption{\label{tab:ulbest}Best 90\% $h_{\rm rss}$ upper limits resulting from
a search of GW signals from GRBs occurring during the three LIGO runs; 25-ms 
crosscorrelation analysis ($\text{Hz}^{-1/2}$)}
\begin{ruledtabular}
\begin{tabular}{cllc}
Run & \multicolumn{1}{c}{$h_{\rm rss,90}$} & \multicolumn{1}{c}{$h_{\rm rss,90}$} & $f_0$           \\
    & \multicolumn{1}{c}{(circular)}       & \multicolumn{1}{c}{(linear)}         &                 \\\hline
    &                                      &                                      &                 \\
S4  & $1.1\times10^{-21}$ (050306)  & $3.6\times10^{-21}$ (050223)  & 150 Hz          \\
S3  & $8.5\times10^{-21}$ (031109A) & $2.9\times10^{-20}$ (031109A) & 250 Hz          \\
S2  & $8.2\times10^{-21}$ (030414)  & $3.1\times10^{-20}$ (030329B) & 250 Hz          \\
\end{tabular}
\end{ruledtabular}
\end{table}

\section{Constraining GRB population models}
\label{popstudy}

The approach of combining multiple GRBs to look for a GW signature associated
with a sample of GRBs was described in Section~\ref{sec:stattest}. Having
established that the null hypothesis is quite significant, i.e., that we cannot
claim the detection of an association between GWs and the GRB population at a
high enough confidence, we turn to setting constraints on the parameters of GRB
population models.  The method is summarized below and
described in detail in \cite{SDM:GWDAW10}.

For a pair of detectors, it can be shown that only three scalar parameters 
associated with a GW signal are sufficient to determine the distribution of 
largest crosscorrelations.  The parameters are the matched filtering 
{\it signal-to-noise ratios} (SNRs) of the strain signals in individual  
detectors and the angle between the two strain signal vectors (as defined by the 
Euclidean inner product).  In the following, a {\em source population model} is 
the joint probability distribution function of these three parameters.

Our approach to putting constraints on source population models follows the
standard frequentist upper limit procedure (cf. Section \ref{sec:grblimits}).  
In this case, let $P(\chi|Z_{\rm source})$ be the marginal cumulative 
probability density function of the sum-max statistic, $\chi$, given the 
population model $Z_{\rm source}$, and let $\chi_\alpha$ be such that 
$P(\chi_\alpha|Z_{\rm source}) = 1-\alpha$, where $0< \alpha< 1$, and $1-\alpha$
is the desired confidence level.  If the {\em observed} value of $\chi$ is 
greater than or equal to $\chi_\alpha$, the corresponding model $Z_{\rm source}$ 
is accepted.  It is rejected when $\chi<\chi_\alpha$.  To obtain the marginal 
distribution of $\chi$, we first construct its conditional distribution for a 
set of $N_{\rm GRB}$ values for the scalar parameters above, where $N_{\rm GRB}$ 
is the number of H1-H2 GRB on-source segments.  The marginal distribution of 
$\chi$ for a given source population model can then be estimated by randomly 
drawing values of the scalar parameters followed by drawing $\chi$ from the 
corresponding conditional distribution.

Since we use only the H1-H2 pair, which are perfectly aligned, the angle between 
the strain responses is zero.  Further, for narrowband signals,the SNR values 
for H1 and H2 can be related by the measurable ratio of their calibrated noise 
power spectral densities (PSDs). Hence, only one  parameter, which we chose to 
be the SNR, $\rho$, of the signal in  H1, is required. Thus, the source population 
model, $Z_{\rm source}$, is simply the univariate distribution of $\rho$.  An 
additional point that needs to be accounted for is the variation in the 
sensitivities of H1 and H2, both within the runs as well as the significant 
improvements from one run to the next.  This is done by fixing a {\em fiducial} 
noise PSD, $S^{(0)} (f)$, and approximating the PSD of H1 for each GRB as simply 
a scaled version of it.  We set the fiducial noise PSD to the one corresponding 
to the initial LIGO design sensitivity for the 4-km 
IFOs\footnote{http://www.ligo.caltech.edu/$\sim$lazz/distribution /LSC\_Data/}
and compute the scale factor at a fixed frequency of $200$~Hz, which was 
approximately where most PSDs had their minimum during the S2, S3, and S4 runs.

We use the theoretical prediction of the observed redshift distribution of GRBs
given in~\cite{bromm+loeb:2002} to construct $Z_{\rm source}$ (prediction for 
the scenario of star formation via atomic hydrogen cooling).  An alternative
is to simply use the measured redshift distribution but \cite{berger05,bagolyi06} 
show that there is a significant selection bias that affects the measured 
redshifts for Swift and non-Swift GRBs, both of which are used in our analysis.  
The model in~\cite{bromm+loeb:2002} is valid for long-duration GRBs, which are 
expected to trace the massive star formation rate of the Universe.  We fit a 
piecewise parabolic curve (with 3 pieces) to figure 1 of \cite{bromm+loeb:2002} 
and then use the same subsequent calculational steps given in
\cite{bromm+loeb:2002} to obtain the redshift distribution for a flux-limited 
detector such as Swift.  Fitting the star formation rate with a smooth curve 
allows us to extend the redshift distribution reliably to very small values of 
the redshift.  Having obtained the distribution, we directly draw random values 
of the redshift, $z$, from it.  Each redshift value is then converted to the 
corresponding luminosity distance $D$ (corresponding to a Friedmann-Robertson-Walker 
cosmological model with $\Omega_m=0.3$, $\Omega_\Lambda = 0.7$ and 
$H_0=72$ km ${\rm sec}^{-1}$ ${\rm Mpc}^{-1}$).

A simple model is used for the GW emission from GRBs.  We assume that GRBs are 
standard candles in GW that emit a fixed amount of energy, $E_{\rm GW}$, 
isotropically with similar amounts of radiation in the two uncorrelated 
polarizations + and $\times$.  Further, neglecting the effect of redshift on the 
signal spectrum, we assume that the spectra of the received signals $h_+$ and 
$h_\times$ are centered at a fixed frequency of $f_o$ in a band that is 
sufficiently narrow such that the noise power spectral density is approximately 
constant over it.  In this case, the SNR is given by
\begin{eqnarray}
\rho & \simeq &\sqrt{2}\,F_{\rm ave}\,\frac{h_{\rm rss}}{\sqrt{S^{(0)}(f_o)}} \;,
\end{eqnarray}
where we have expressed the SNR with respect to the fiducial noise PSD.  Since 
the emission is isotropic, the energy emitted in gravitational waves is 
(cf. Section\linebreak VIIIA),
\begin{equation}
E_{\rm GW} \approx \frac{\pi^2 c^3}{G}\,\frac{D^2}{1+z}\, f_o^2 h_{\rm rss}^2 \;.
\label{egw_hrss}
\end{equation}
To convert the luminosity distance, $D$, for a given GRB into SNR $\rho$, we use 
the normalization
\begin{equation}
\rho = \sqrt{2}\,F_{\rm ave}\,\rho_0\,\frac{D_0}{D}\left(\frac{1+z}{1+z_0}\right)^{3/2}\;,
\end{equation}
where $D_0$ is chosen to be the most probable luminosity distance, at the 
corresponding redshift $z_0$, and $\rho_0$ is the observed SNR for a GRB that 
occured at $D_0$ with an optimal sky location and the above properties for 
$h_+$, $h_\times$ and $E_{\rm GW}$.  The redshift distribution predicted 
in~\cite{bromm+loeb:2002} for Swift has a peak at $z = 1.8$, which yields 
$D_0 = 13.286$~Gpc.  The acceptance-rejection rule above simply becomes an upper 
limit on the value of $\rho_0$.  Note that, because of the scaling of noise PSDs 
discussed above, $\rho_0$ should be understood as the SNR of the strain response 
(for a GRB directly above the detector) that operates at design sensitivity.  
For GRBs that do not have direction information, random values for $F_{\rm ave}$ 
are drawn from a uniform distribution on the celestial sphere.

Finally, in terms of the upper limit, $\rho_{\rm upper}$, obtained on $\rho_0$, 
we get an upper limit on $E_{\rm GW}$,
\begin{equation}
E_{\rm GW} \leq  \frac{\pi^2 c^3}{G}     
           \frac{D_0^2}{1+z_0}f_o^2 S^{(0)}(f_o)\rho_{\rm upper}^2\;.
\label{egw_upper}
\end{equation}
For $z_0=1.8$, $f_o=200$~Hz, and
$\sqrt{S^{(0)}(f_o)}=2.98\times 10^{-23}\,\text{Hz}^{-1/2}$, we get
$E_{\rm GW} \leq 8.43\times 10^{55} \rho_{\rm upper}^2$~ergs
($\equiv~47.3\rho_{\rm upper}^2$~$M_\odot c^2$).

\begin{figure}
\includegraphics[width=3.70in]{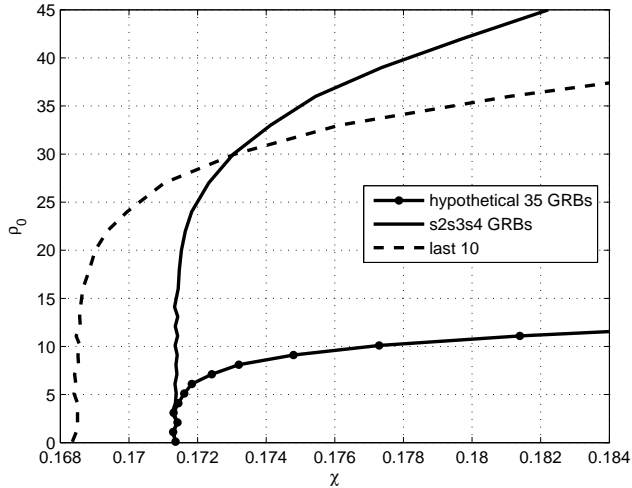}
\caption{Upper limit confidence belts at 90\% confidence level on $\rho_0$, the 
SNR at the most probable redshift for Swift GRBs given in \cite{bromm+loeb:2002}.  
The solid line is the curve for all S2, S3, S4 GRBs that were used in the H1-H2 
search (on-source $\chi = 0.1753$).  The dashed line is the curve when only the 
last 10 GRBs from the above set are selected (on-source $\chi = 0.1702$).  The 
line with filled circles is for a hypothetical scenario with 35 GRBs, all with an 
optimal sky location, and two identical and constant sensitivity detectors.  The 
shifting of the curves horizontally is due to the change in the variance of 
$\chi$ as the number of GRBs is changed.  For each value of $\rho_0$,  10,000 
values of $\chi$ were drawn from its marginal distribution.
\label{confbelt_bl2002_s2s3s4}}
\end{figure}

Figure~\ref{confbelt_bl2002_s2s3s4} shows the 90\% upper limit confidence belt for 
$\rho_0$.  The on-source value of sum-max was $\chi=0.1753$ for the S2, S3, S4 
GRB sample.  Hence, $\rho_0\leq 35.5$ and 
$E_{\rm GW} \leq 5.96\times 10^4$~$M_\odot c^2$.  This limit is too high to be 
of any astrophysical importance. However, as discussed later, future analyses 
may be able to improve by orders of magnitude on this result.

Since the detectors during the S2 run were much less sensitive than S4, one may 
expect that dropping the S2 GRBs from the analysis can improve the upper limit. 
Figure~\ref{confbelt_bl2002_s2s3s4} shows the 90\% level upper limit belt obtained 
for the case when only the last 10 GRBs, spanning the whole of S4 and part of 
S3, were retained in the analysis.  The corresponding value of $\chi = 0.1702$ 
yields an upper limit of $24.6$ on $\rho_0$.  Thus, we obtain
$E_{\rm GW} \leq 2.86\times 10^4$~$M_\odot c^2$.  This shows, as expected, that
making judicious cuts on the sample of GRBs can lead to improvements in upper 
limits.  The upper limit can probably be improved further by retaining only the 
S4 GRBs, but for a small number of GRBs the distribution of $\chi$ used is not 
valid and a more accurate calculation has to be done.  In 
Fig.~\ref{confbelt_bl2002_s2s3s4}, we also show the upper limit confidence belt 
for a hypothetical scenario that is likely for the ongoing S5 run: a sample size 
of about 35 GRBs with the H1 and L1 detectors operating at design sensitivity. 
The implications of this curve are discussed in the next section.

The confidence belt construction outlined in this section is for illustrative 
purposes only.  In particular, we have not taken into account factors such as 
(i) changing noise spectral shapes, (ii) red-shifting of the standard candle 
(K-correction) and possible systematic errors associated with the population 
model used.  A more comprehensive investigation is planned for the S5 data.

\section{\label{sec:discuss}Discussion}

This search is not very restrictive with respect to models for astrophysical 
systems which give rise to GRBs.  The main assumption we have made is that the 
GW emission is limited in duration --- we sum over periods of up to 100 ms, 
which is much greater than the characteristic times expected for GW burst emission 
in most GRB models.  Given the LIGO sensitivity at the time of this search, it is 
not surprising that our experimental limits in this search do not place 
significant restrictions on the astrophysical models at present.  However, given 
the rapid development of the field, it is not precluded that the limits presented 
here will provide guidance to GRB astrophysics in the near future.  In any case, 
it is useful to get a sense for the interplay between the measured 
gravitational-wave strain limits for individual GRBs from 
Section~\ref{sec:grblimits} and astrophysical models. So in this section we 
provide some astrophysical context to our experimental limits.  We emphasize 
that the estimates given below are for illustration, and are not to be construed 
as measured astrophysical limits.

The local gravitational-wave energy flux in the two independent polarizations, 
$h_+(t)$ and $h_\times(t)$, is \cite{300years,isaacson}
\begin{equation}
\frac{dE}{dA\,dt}= \frac{1}{16\pi}\,\frac{c^3}{G}\left(\dot{h}_+^2 + \dot{h}_\times^2\right)
\label{eq:eflux}
\end{equation}
which can be integrated over the duration of a burst of gravitational radiation
and over a closed surface to relate the strains evaluated on the surface to the 
total intrinsic energy associated with a source within this volume.  For a 
source at the center of a sphere of radius $r$ at negligible redshift, then 
$dA = r^2 d\Omega$, as usual.

Since many of the GRBs in the sample are found to have significant redshifts, it 
is useful to generalize the above to cosmological distances.  In this case, we 
can use the luminosity distance, $D$, which by definition relates the intrinsic 
luminosity, $\cal L$, of an isotropically emitting source to the energy flux 
$\cal F$ at a detector by ${\cal L}= {\cal F}\,(4\pi D^2)$.  For a non-isotropic 
emitter, we replace the $4\pi$ by an integration over solid angle.  We note that 
$\cal F$ is by definition the left hand side of Eq.~\ref{eq:eflux}, and the 
intrinsic luminosity is ${\cal L}=dE_e/dt_e$.  To integrate this over the signal 
duration at the detector, we use $dt = (1+z) dt_e$.  Hence, the energy emitted in 
gravitational radiation is,
\begin{align}
E_e & = \frac{D^2}{1+z} \; \int d\Omega \int {\cal F}\> dt \nonumber \\
    & = \frac{1}{16\pi}\,\frac{c^3}{G}\; \frac{D^2}{1+z}\,\int d\Omega 
        \int\left(\dot{h}_+^2 + \dot{h}_\times^2\right)\> dt
\label{eq:EGWz}
\end{align}
For negligible redshifts, $D=r$. We note that $D=D(z)$ is itself a function
of the redshift, and in general depends on the cosmological model.

If the signal power at the detectors is dominated by a frequency $f_o$, as is 
the case for the sine-gaussian waveforms introduced earlier, then Eq.~\ref{eq:EGWz}
can be written in the approximate form
\begin{equation}
  E_e\approx\frac{\pi}{4}\,\frac{c^3}{G}\,\frac{D^2}{1+z}\>f_o^2 
  \int d\Omega \,\int \left(h_+^2 + h_\times^2\right)\,dt ~~~,
\label{eq:Esimp}\end{equation}
which allows a direct relation between $E_e$ and the observable $h_{\rm rss}$
(see Eqs.~\ref{eq:rssdef1} and \ref{eq:rssdef2}).  For sine-gaussian waveforms, the 
approximation is quite good for $Q \gtrsim 3$; the error is approximately $1/(1+2Q^2)$.  
We will assume here that the simulated waveforms are effectively local to the detectors. 
Specifically, the frequency $f_o$ is the measured frequency (which is related to 
the source frequency $f_e$ by $f_o=f_e/(1+z)$).  Of course, some fraction of the 
source power might be shifted in or out of the sensitive LIGO band in frequency 
or expanded in time beyond our integration time.  We ignore any such effect here.

\subsection{Case I: Isotropic emission}

First, we consider a simple, but unphysical, example where the radiation is 
emitted isotropically, with equal power in the (uncorrelated) $+$ and $\times$ 
polarizations.  In this case, Eq.~\ref{eq:Esimp} becomes
\begin{equation}
E_{\rm iso}\approx \frac{\pi^2 c^3}{G}\,\frac{D^2}{1+z}\, f_o^2 h_{\rm rss}^2 ~~~.
\label{eq:isoopt}\end{equation}
Then for a LIGO sensitivity for some waveform represented by $h_{\rm rss}$, we 
might hope to be sensitive to a distance
\begin{eqnarray}
D \approx 70 \>{\rm Mpc}\left(\frac{100\> {\rm Hz}}{f_o}\right)\,
\left(\frac{E_{\rm iso}}{M_\odot c^2}\right)^{1/2}\, \nonumber \\\times
\left(\frac{10^{-21}\>{\rm Hz}^{-1/2}}{h_{\rm rss}}\right)\,(1+z)^{1/2}
\label{eq:Diso}
\end{eqnarray}
for an isotropic source which emits gravitational-wave energy $E_{\rm iso}$ (in 
units of solar rest energy) at detected frequency $f_o$.

\subsection{Case II: Long-duration GRBs}

For long-duration GRBs, we consider the scenario where such events are 
associated with a core collapse, perhaps involving a very massive progenitor 
\cite{LGRBvsSN}.  Gravitational wave emission has been simulated for supernova 
core collapse models for relatively light ($\sim 10 M_\odot$) progenitors, for 
example, in Refs.~\cite{ottburrows04,dfm02}.  These models invoke axisymmetry, 
with linearly polarized strain that is proportional to $\sin^2\iota$, where 
$\iota$ is the angle with respect to the symmetry axis.

Integrating over the full solid angle, Eq.~\ref{eq:Esimp} becomes
\begin{equation}
E_{\rm sn}\approx \frac{8\pi^2 c^3}{15G}\,\frac{D^2}{1+z}\,\frac{f_o^2 \> 
h_{\rm rss}^2}{\sin^4\!\iota} ~~~~.
\label{eq:EGWlong}\end{equation}
We then find an analogous expression to Eq.~\ref{eq:Diso},
\begin{eqnarray}
D \approx 1 \>{\rm Mpc}\left(\frac{100\> {\rm Hz}}{f_o}\right)\,
\left(\frac{E_{\rm sn}}{10^{-4}M_\odot c^2}\right)^{1/2}\, \nonumber \\\times
\left(\frac{10^{-21}\>{\rm Hz}^{-1/2}}{h_{\rm rss}}\right)\, \sin^2\!\iota\>(1+z)^{1/2}
\label{eq:Dsn}
\end{eqnarray}
As described earlier, our experimental limits correctly account for the antenna
pattern associated with each GRB.  Hence, no additional factors are required in 
the equation above if one were to use values from the tables of results.  
However, if one wished, for example, to apply a theoretical $h_{\rm rss}$ to a
particular GRB, the antenna factors for each GRB are given in 
Table~\ref{tab:GRBsample}.

Core collapse simulations indicate that most of the gravitational radiation is 
emitted from the core bounce, and that $E_{\rm sn}$ should be at most 
$10^{-7}M c^2$ \cite{dfm02}, or even smaller \cite{ottburrows04}.  For the very 
massive progenitors often associated with long-duration GRBs, the collapse 
process is uncertain.  Whether there is a bounce at all, or simply a direct 
collapse to a black hole, depends \cite{Fryer} on the mass, metallicity, and 
angular momentum of the progenitor.  In any case, there is no reason to believe 
that the efficiency for converting the collapse into gravitational radiation 
increases with the progenitor mass.

In fact, the situation for GW detection in this scenario is especially 
unpromising.  It is natural to align the symmetry axis of the (rotating) core 
collapse with the direction of the gamma-ray beam. Hence, $\iota=0$ would be 
along the line of sight to the detectors. For a typical gamma-ray beaming angle 
of half-width $\sim 10^\circ$, then at best, where the detectors are at the edge
of the beam, this would give a suppression factor of $\sim 30$.  Finally, we 
note that long-duration GRBs are distant objects, with mean observed redshift of
$\approx 2.4$.\footnote{http://swift.gsfc.nasa.gov}  Given their redshift 
distribution, the simulations to date indicate that detection of long-duration 
GRBs is unlikely if core bounce is the dominant radiation mechanism.

However, core collapse can potentially drive other mechanisms more favorable for 
gravitational radiation detection. In particular, bar mode instabilities are 
potentially very efficient radiators and do not suffer from the unfavorable 
alignment noted above for axisymmetric core bounces.  Similarly, core 
fragmentation during collapse can lead to GW radiation from the inspiraling 
fragments.  Reference \cite{Fryer} has examined these possibilities, and while 
the likelihood of bar instabilities or core fragmentation, along with their
detailed properties, is uncertain, the resulting gravitational radiation is 
plausibly detectable for a nearby GRB.  In such cases, Eqs.~\ref{eq:Emerge} 
and \ref{eq:Dmerge} might be more appropriate descriptions of the radiated 
energy and distance to which we can detect the source.

The nearest known GRB to date is long-duration burst GRB~980425 at $D=35$ Mpc.  
From Eq.~\ref{eq:Dsn}, LIGO detection at 35 Mpc by the method described in this 
paper would require an efficiency of at least 
$E_{\rm sn}/M_\odot c^2 \approx 10\%$ for a $1M_\odot$ system, much larger than 
the efficiency expected from conventional core collapse, but perhaps not 
unreasonable in case of bar instabilities or core fragmentation.  Unfortunately, 
the data considered here did not include any such nearby events.  For example, 
during the (most sensitive) S4 run, the GRB sample consisted of only 4 events, 
all long-duration GRBs.  The most nearby of these with a measured redshift was 
GRB~050223 ($z=0.5915$) at $D\approx 3.5$ Gpc. Assuming linear polarization, we 
can obtain an estimate for sensitivity from the 90\% upper limit for GRB~050223 
from Table~\ref{tab:ulbest}.  This gives for $E_{\rm sn}$ the value
$1.6\times 10^4\,M_\odot c^2$.  This is in fact very close to the source 
luminosity maximum of $c^5/G$ \cite{dyson}, which gives $2\times 10^4\,M_\odot c^2$ 
if sustained for 100 ms. The larger sample of GRBs in future runs will hopefully
include some long-duration GRBs at smaller redshift.

\subsection{Case III: Short-duration GRBs}

Short-duration GRBs, to the extent that the population is associated with the 
merger of compact binary systems, offer several potentially interesting
characteristics.  First, such mergers are found to be relatively efficient
radiators of gravitational radiation.  Second, the emission pattern is not 
expected to be problematic.  Moreover, the measured redshifts to date indicate a
significant number of relatively low-$z$ GRBs.  (The average redshift was
$\approx 0.4$ for the 2005 sample of 5 events.)  The mergers may include 
formation of a hypermassive neutron star \cite{HMNS} or a black hole with 
associated ringdown \cite{NSNSmerge2}.  Finally, the merger GW emission, which 
is best suited to the methodology described in this paper, would be preceded by 
an inspiral which is potentially detectable by a sensitive, independent LIGO 
search based on matching inspiral waveform templates \cite{LIGOinspiral}.  
However, we have verified that the present search, while not as sensitive to
inspirals as the dedicated waveform template-based search, can readily detect 
inspiral emission when there is sufficient signal to background in individual 
25-ms or 100-ms bins.  In this case, the maximum crosscorrelation occurs when 
the frequency of the inspiral radiation passes through the 100-300 Hz range, 
where the detector sensitivity is best (see Fig.~\ref{fig:ligoruns}).  Therefore, 
while this search is uniquely sensitive to the higher frequency, short-duration, 
poorly modeled gravitational waves from the merger phase, it also provides 
independent information on the inspiral phase.  Recent estimates 
\cite{NYF,guetta06} place the chance for detection of a BH-NS merger at up to 
$\sim30$\% for a year of simultaneous LIGO and Swift operation, and $\sim10$\% 
for a NS-NS merger.  Here, we provide an estimate for a contrived, but physically 
motivated, model.

We suppose that the gravitational-wave emission pattern for the merger follows 
that of the inspiral, that is
\begin{equation}
h_+ = h_o\,f(t)\> \frac{1}{2}(1+\cos^2\iota)\>; 
~~~~ h_\times = h_o\,g(t)\> \cos\iota
\label{eq:hmerg}\end{equation}
where $\iota$ is measured with respect to the axis orthogonal to the plane of 
the inspiral orbit.  The functions $f(t)$ and $g(t)$ are orthogonal functions, 
for example $f(t)$ could be the sine-gaussian form discussed earlier, while 
$g(t)$ is a cosine-gaussian; $h_o$ represents a constant amplitude.  While the 
degree of gamma-ray beaming for short-duration GRBs is still uncertain, we 
suppose that the gamma rays are preferentially emitted along the angular 
momentum axis of the merger system. If the Earth is near the center of the 
gamma-ray beam, then $\iota=0$ is along the line of sight between detector and 
source, which is a maximum of the assumed emission pattern, and the radiation 
will be circularly polarized.  Returning to Eq.~\ref{eq:Esimp} and integrating 
over the full solid angle, we find in this case (with $\iota=0$):
\begin{equation}
E_{\rm merge} \approx \frac{2\pi^2}{5}\,\frac{c^3}{G}\,
\frac{D^2}{1+z}\>\ f_o^2 \, \>h_{\rm rss}^2  ~~~.
\label{eq:Emerge}\end{equation}
Rewriting this for $D$, as before, gives
\begin{eqnarray}
D\approx 44 \>{\rm Mpc}\left(\frac{250\> {\rm Hz}}{f_o}\right)\,
\left(\frac{E_{\rm merge}}{M_\odot c^2}\right)^{1/2}\, \nonumber \\\times
\left(\frac{10^{-21}\>{\rm Hz}^{-1/2}}{h_{\rm rss}}\right)\,(1+z)^{1/2} ~~~.
\label{eq:Dmerge}
\end{eqnarray}
The comments below Eq.~\ref{eq:Dsn} concerning antenna factors also apply here.

There has been substantial recent progress in calculations of gravitational 
radiation production in various types of mergers.  Numerical simulations of 
NS-NS mergers give \cite{NSNSmerge,NSNSmerge2,oechslin07} typical values of the 
radiated energy of about 0.5-1\% of the total mass, or $E\approx 0.01 M_\odot c^2$.  
These simulations indicate that the frequency spectrum can be broad, ranging 
from a few hundred Hz to $\sim 2$~kHz.  Perhaps the most interesting case is 
BH-NS mergers.  Very recent calculations \cite{BHNSmerge1,BHNSmerge2,faber06} 
indicate radiative energies ranging from $\sim 10^{-4}$ to $\sim 10^{-2}$ of the 
total mass, where the range is likely to reflect the very different initial 
conditions assumed in the simulations.  While there are no short-duration GRBs 
in the S4 sample, we can use typical upper limits on $h_{\rm rss}$ from Table 
\ref{tab:s4ul25mspsic} as an indication of sensitivity.  For example a 
$1.4 M_\odot$ NS plus $10 M_\odot$ BH binary system would have merger GW emission 
at frequencies starting at about 400 Hz.  If this system were to radiate 1\% of 
its rest energy into gravitational radiation at 400 Hz, the distance sensitivity 
would be $D\sim 5$~Mpc.  The search would also be sensitive to the inspiral 
emission from this system at lower ($\sim 200$ Hz) frequency.

\subsection{Prospects}

Here we discuss the future prospects for science run S5 and beyond.  At the 
sensitivity for science run S4, the prospects for detection are clearly 
dominated by the possibility of a nearby GRB.  While this distance scale is 
guided by the discussion above, we are prepared to be surprised by new 
mechanisms for GW emission.  Nevertheless, we expect detection of individual 
GRBs to depend in no small part on the appearance of a ``special'' event.  Thus, 
a data sample which includes a large number of GRBs is especially important. 
For science run S5, the LIGO detectors will be operating at design 
sensitivity and fully coincident with Swift operation.  This should yield over 
100 GRBs, including some with redshift measurements.  And clearly, the search 
radius will increase in proportion to improvements in the LIGO strain 
sensitivity.

The results pertaining to the GRB population obtained in Section \ref{popstudy} 
will certainly improve for the S5 run and in future observations with Advanced 
LIGO.  To make an estimate, we look at the various factors involved in 
Eq.~\ref{egw_upper} for the upper limit on $E_{\rm GW}$.  As a reference, we use 
the limit obtained here using all S2, S3 and S4 GRBs.  Since most factors in
Eq.~\ref{egw_upper} come as squares, moderate improvements in each has a 
significant overall effect.

Since the direction to each GRB will be known, it may be possible to select a 
subsample of, say, 35 GRBs from the sample in S5 (i.e., about the same number as 
the whole of S2, S3 and S4) such that $\bigl<F_{\rm ave}\bigr> \simeq 1/\sqrt{2}$, 
the maximum possible.  Further, assume that we use H1-L1 crosscorrelations.
Figure~\ref{confbelt_bl2002_s2s3s4} shows the confidence belt for the case of 35 
optimally located GRBs and a pair of identical detectors.  One can expect to get 
an upper limit of $\simeq 10$  on $\rho_0$ with this curve, which is a factor of 
$\sim 3.5$ better than the current limit on $\rho_0$.

Without altering other parameters of the analysis, therefore, we can expect 
$3.5^2$ or, in round numbers, a factor of $\sim 10$ improvement in the upper
limit on $E_{\rm GW}$ for S5.  Additional improvements are possible by imposing 
a cut based on measured redshifts, in addition to the cut on sky positions, and 
by reducing the search interval from the current value of 180~seconds.  Looking 
beyond S5, the most obvious source of improvement would be the $\geq 10$ factor 
of improvement in the strain noise level when Advanced LIGO comes online around 
the middle of the next decade.  This translates into an additional factor of 
$\geq 100$ reduction in the upper limit.  When Advanced LIGO comes online, there 
may be a worldwide network of GW detectors of comparable sensitivity.  Besides 
allowing a more uniform sky coverage, resulting in a larger sample of GRBs with 
optimal orientation, network analysis methods 
\cite{klimenko+etal:05,Rakhmanov:gwdaw10,mohanty+etal:06} that make more optimal 
use of data from multiple detectors can be used to increase the base sensitivity 
of the method.  Finally, with enough GRBs, we could separately analyze the class 
of long and short duration bursts.  Since the most probable redshift for 
short-duration GRBs is expected to be inherently smaller, we could obtain 
significantly tighter constraints on the energy emitted in gravitational waves 
from this class of GRBs.

The discussion above was confined to a particular model for GRB redshift 
distribution and GW emission.  Further work is needed to develop more general 
analysis methods that can be applied to a wider variety of models and that take 
better account of prior information from existing observations.

\section{Summary and conclusion}

We searched for gravitational-wave bursts, targetting short GW signals with 
durations from $\sim$~1~ms to $\sim$~100~ms, associated with 39 GRBs that were
detected by gamma-ray satellite experiments while the S2, S3, and S4 science
runs of the LIGO experiment were in progress.  To take into account the unknown
onset time of the GW signal relative to the GRB trigger time, the search covered
180 seconds of data surrounding the GRB trigger times.  These 180-second data 
segments from the different IFOs were crosscorrelated to probe for correlated 
signals.  We searched for an association on an individual-GRB basis, and also 
applied different statistical tests to search for the cumulative effect of weak 
GW signals.  We found no evidence for gravitational-wave burst emission 
associated with the GRB sample examined using the different search methods.

Using simulated Q = 8.9 sine-gaussian waveforms and the direction-dependent
antenna response of the interferometers to a GW source, we obtained upper limits
on the root-sum-square amplitude of linearly polarized and circularly polarized
gravitational waves from each of 22 GRBs with well-localized positions.  
Associating these limits with the energy radiated by the GRB sources into 
gravitational radiation is inherently speculative at this stage of development 
of the field and depends crucially on the astrophysical scenario one adopts for 
the GRB progenitors. The most favorable cases considered here suggest that the 
LIGO sensitivity for run S4 would allow sensitivity to a solar mass-equivalent 
of radiated GW energy to distances of tens of Mpc.

The sample of GRBs was combined to set an upper limit on the GW energy emitted 
using a simple standard candle model and a theoretical redshift distribution of 
GRBs.  Although the upper limit obtained is not astrophysically important, a 
straightforward and realistic extrapolation to future observations suggests that 
this limit can be improved by orders of magnitude.  It may be possible to set a
sub-solar mass limit when Advanced LIGO comes online. This would put us in an
astrophysically interesting regime since at least one model~\cite{vanPutten:grb} 
predicts an energy loss of 0.2 solar masses for long-duration GRBs.

It is opportune that Swift will be operating and detecting GRBs at the time
when the fifth science run of LIGO, S5, will be in progress.  The goal for the
S5 run is to collect one year of coincident LHO-LLO data at the design
sensitivity.  Given the Swift GRB detection rate, we anticipate an S5 sample of
more than 100 GRB triggers that can be used to further probe for gravitational
radiation associated with GRBs.  It is hoped that a large GRB sample will
increase the chances for finding such an association.

\begin{acknowledgments}
The authors gratefully acknowledge the support of the United States National 
Science Foundation for the construction and operation of the LIGO Laboratory 
and the Particle Physics and Astronomy Research Council of the United Kingdom, 
the Max-Planck-Society and the State of Niedersachsen/Germany for support of 
the construction and operation of the GEO600 detector. The authors also 
gratefully acknowledge the support of the research by these agencies and by the 
Australian Research Council, the Natural Sciences and Engineering Research 
Council of Canada, the Council of Scientific and Industrial Research of India, 
the Department of Science and Technology of India, the Spanish Ministerio de 
Educacion y Ciencia, The National Aeronautics and Space Administration, 
the John Simon Guggenheim Foundation, the Alexander von Humboldt Foundation, 
the Leverhulme Trust, the David and Lucile Packard Foundation, 
the Research Corporation, and the Alfred P. Sloan Foundation.
This document has been assigned LIGO Laboratory document number 
LIGO-P060024-07-Z.
\end{acknowledgments}

\bibliography{multigrb_prd}

\begin{thebibliography}{76}
\expandafter\ifx\csname natexlab\endcsname\relax\def\natexlab#1{#1}\fi
\expandafter\ifx\csname bibnamefont\endcsname\relax
  \def\bibnamefont#1{#1}\fi
\expandafter\ifx\csname bibfnamefont\endcsname\relax
  \def\bibfnamefont#1{#1}\fi
\expandafter\ifx\csname citenamefont\endcsname\relax
  \def\citenamefont#1{#1}\fi
\expandafter\ifx\csname url\endcsname\relax
  \def\url#1{\texttt{#1}}\fi
\expandafter\ifx\csname urlprefix\endcsname\relax\def\urlprefix{URL }\fi
\providecommand{\bibinfo}[2]{#2}
\providecommand{\eprint}[2][]{\url{#2}}

\bibitem[{\citenamefont{Klebesadel et~al.}(1973)\citenamefont{Klebesadel,
  Strong, and Olson}}]{klebesadel73}
\bibinfo{author}{\bibfnamefont{R.~W.} \bibnamefont{Klebesadel}},
  \bibinfo{author}{\bibfnamefont{I.~B.} \bibnamefont{Strong}},
  \bibnamefont{and} \bibinfo{author}{\bibfnamefont{R.~A.} \bibnamefont{Olson}},
  \bibinfo{journal}{Astrophys. J.} \textbf{\bibinfo{volume}{182}},
  \bibinfo{pages}{L85} (\bibinfo{year}{1973}).

\bibitem[{\citenamefont{Fishman et~al.}(1992)\citenamefont{Fishman, Meegan,
  Wilson, Paciesas, and Pendleton}}]{fishman92}
\bibinfo{author}{\bibfnamefont{G.~J.} \bibnamefont{Fishman}},
  \bibinfo{author}{\bibfnamefont{C.~A.} \bibnamefont{Meegan}},
  \bibinfo{author}{\bibfnamefont{R.~B.} \bibnamefont{Wilson}},
  \bibinfo{author}{\bibfnamefont{W.~S.} \bibnamefont{Paciesas}},
  \bibnamefont{and} \bibinfo{author}{\bibfnamefont{G.~N.}
  \bibnamefont{Pendleton}}, in \emph{\bibinfo{booktitle}{The Compton
  Observatory Science Workshop}} (\bibinfo{year}{1992}), pp.
  \bibinfo{pages}{26--34}.

\bibitem[{\citenamefont{Boella et~al.}(1997)}]{boella97}
\bibinfo{author}{\bibfnamefont{G.}~\bibnamefont{Boella}} \bibnamefont{et~al.},
  \bibinfo{journal}{Astron. \& Astrophys.} \textbf{\bibinfo{volume}{122}},
  \bibinfo{pages}{299} (\bibinfo{year}{1997}).

\bibitem[{\citenamefont{Meegan et~al.}(1992)}]{meegan92}
\bibinfo{author}{\bibfnamefont{C.~A.} \bibnamefont{Meegan}}
  \bibnamefont{et~al.}, \bibinfo{journal}{Nature}
  \textbf{\bibinfo{volume}{355}}, \bibinfo{pages}{143} (\bibinfo{year}{1992}).

\bibitem[{\citenamefont{Kouvelitou et~al.}(1993)}]{ck93}
\bibinfo{author}{\bibfnamefont{C.}~\bibnamefont{Kouvelitou}}
  \bibnamefont{et~al.}, \bibinfo{journal}{Astrophys. J.}
  \textbf{\bibinfo{volume}{413}}, \bibinfo{pages}{L101} (\bibinfo{year}{1993}).

\bibitem[{\citenamefont{Costa et~al.}(1997)}]{costa97}
\bibinfo{author}{\bibfnamefont{E.}~\bibnamefont{Costa}} \bibnamefont{et~al.},
  \bibinfo{journal}{Nature} \textbf{\bibinfo{volume}{387}},
  \bibinfo{pages}{783} (\bibinfo{year}{1997}).

\bibitem[{\citenamefont{van Paradijs et~al.}(1997)}]{paradijs97}
\bibinfo{author}{\bibfnamefont{J.}~\bibnamefont{van Paradijs}}
  \bibnamefont{et~al.}, \bibinfo{journal}{Nature}
  \textbf{\bibinfo{volume}{386}}, \bibinfo{pages}{686} (\bibinfo{year}{1997}).

\bibitem[{\citenamefont{Frail et~al.}(1997)\citenamefont{Frail, Kulkarni,
  Nicastro, Feroci, and Taylor}}]{frail97}
\bibinfo{author}{\bibfnamefont{D.~A.} \bibnamefont{Frail}},
  \bibinfo{author}{\bibfnamefont{S.~R.} \bibnamefont{Kulkarni}},
  \bibinfo{author}{\bibfnamefont{L.}~\bibnamefont{Nicastro}},
  \bibinfo{author}{\bibfnamefont{M.}~\bibnamefont{Feroci}}, \bibnamefont{and}
  \bibinfo{author}{\bibfnamefont{G.~B.} \bibnamefont{Taylor}},
  \bibinfo{journal}{Nature} \textbf{\bibinfo{volume}{389}},
  \bibinfo{pages}{261} (\bibinfo{year}{1997}).

\bibitem[{\citenamefont{Metzger et~al.}(1997{\natexlab{a}})}]{metzger97}
\bibinfo{author}{\bibfnamefont{M.~R.} \bibnamefont{Metzger}}
  \bibnamefont{et~al.}, \bibinfo{journal}{IAU Circular}
  \textbf{\bibinfo{volume}{6655}} (\bibinfo{year}{1997}{\natexlab{a}}).

\bibitem[{\citenamefont{Metzger et~al.}(1997{\natexlab{b}})}]{cohen97}
\bibinfo{author}{\bibfnamefont{M.~R.} \bibnamefont{Metzger}}
  \bibnamefont{et~al.}, \bibinfo{journal}{IAU Circular}
  \textbf{\bibinfo{volume}{6676}} (\bibinfo{year}{1997}{\natexlab{b}}).

\bibitem[{\citenamefont{Bloom et~al.}(1998)\citenamefont{Bloom, Djorgovski, and
  Kulkarni}}]{bloom98}
\bibinfo{author}{\bibfnamefont{J.~S.} \bibnamefont{Bloom}},
  \bibinfo{author}{\bibfnamefont{S.~G.} \bibnamefont{Djorgovski}},
  \bibnamefont{and} \bibinfo{author}{\bibfnamefont{S.~R.}
  \bibnamefont{Kulkarni}}, \bibinfo{journal}{Astrophys. J.}
  \textbf{\bibinfo{volume}{507}}, \bibinfo{pages}{L25} (\bibinfo{year}{1998}).

\bibitem[{\citenamefont{Galama et~al.}(1998)}]{galama98}
\bibinfo{author}{\bibfnamefont{T.~J.} \bibnamefont{Galama}}
  \bibnamefont{et~al.}, \bibinfo{journal}{Nature}
  \textbf{\bibinfo{volume}{395}}, \bibinfo{pages}{670} (\bibinfo{year}{1998}).

\bibitem[{\citenamefont{Iwamoto et~al.}(1998)}]{iwamoto98}
\bibinfo{author}{\bibfnamefont{K.}~\bibnamefont{Iwamoto}} \bibnamefont{et~al.},
  \bibinfo{journal}{Nature} \textbf{\bibinfo{volume}{395}},
  \bibinfo{pages}{672} (\bibinfo{year}{1998}).

\bibitem[{\citenamefont{Kulkarni et~al.}(1998)}]{kulkarni98}
\bibinfo{author}{\bibfnamefont{S.~R.} \bibnamefont{Kulkarni}}
  \bibnamefont{et~al.}, \bibinfo{journal}{Nature}
  \textbf{\bibinfo{volume}{395}}, \bibinfo{pages}{663} (\bibinfo{year}{1998}).

\bibitem[{\citenamefont{Gehrels et~al.}(2004)}]{swift04}
\bibinfo{author}{\bibfnamefont{N.}~\bibnamefont{Gehrels}} \bibnamefont{et~al.},
  \bibinfo{journal}{Astrophys. J.} \textbf{\bibinfo{volume}{611}},
  \bibinfo{pages}{1005} (\bibinfo{year}{2004}).

\bibitem[{\citenamefont{Ricker et~al.}(2003)}]{hete03}
\bibinfo{author}{\bibfnamefont{G.~R.} \bibnamefont{Ricker}}
  \bibnamefont{et~al.}, in \emph{\bibinfo{booktitle}{AIP Conf. Proc. 662:
  Gamma-Ray Burst and Afterglow Astronomy 2001: A Workshop Celebrating the
  First Year of the HETE Mission}} (\bibinfo{year}{2003}), pp.
  \bibinfo{pages}{3--16}.

\bibitem[{\citenamefont{Winkler et~al.}(2003)}]{integral03}
\bibinfo{author}{\bibfnamefont{C.}~\bibnamefont{Winkler}} \bibnamefont{et~al.},
  \bibinfo{journal}{Astron. \& Astrophys.} \textbf{\bibinfo{volume}{411}},
  \bibinfo{pages}{L1} (\bibinfo{year}{2003}).

\bibitem[{\citenamefont{Woosley}(1993)}]{woosley93}
\bibinfo{author}{\bibfnamefont{S.~E.} \bibnamefont{Woosley}},
  \bibinfo{journal}{Astrophys. J.} \textbf{\bibinfo{volume}{405}},
  \bibinfo{pages}{273} (\bibinfo{year}{1993}).

\bibitem[{\citenamefont{Eichler et~al.}(1989)\citenamefont{Eichler, Livio,
  Piran, and Schramm}}]{schramm89}
\bibinfo{author}{\bibfnamefont{D.}~\bibnamefont{Eichler}},
  \bibinfo{author}{\bibfnamefont{M.}~\bibnamefont{Livio}},
  \bibinfo{author}{\bibfnamefont{T.}~\bibnamefont{Piran}}, \bibnamefont{and}
  \bibinfo{author}{\bibfnamefont{D.~N.} \bibnamefont{Schramm}},
  \bibinfo{journal}{Nature} \textbf{\bibinfo{volume}{340}},
  \bibinfo{pages}{126} (\bibinfo{year}{1989}).

\bibitem[{\citenamefont{Paczy{\'{n}}ski}(1991)}]{paczynski91}
\bibinfo{author}{\bibfnamefont{B.}~\bibnamefont{Paczy{\'{n}}ski}},
  \bibinfo{journal}{Acta Astron.} \textbf{\bibinfo{volume}{41}},
  \bibinfo{pages}{257} (\bibinfo{year}{1991}).

\bibitem[{\citenamefont{Hjorth et~al.}(2003)}]{hjorth03}
\bibinfo{author}{\bibfnamefont{J.}~\bibnamefont{Hjorth}} \bibnamefont{et~al.},
  \bibinfo{journal}{Nature} \textbf{\bibinfo{volume}{423}},
  \bibinfo{pages}{847} (\bibinfo{year}{2003}).

\bibitem[{\citenamefont{Woosley et~al.}(1999)\citenamefont{Woosley, Eastman,
  and Schmidt}}]{woosley99}
\bibinfo{author}{\bibfnamefont{S.~E.} \bibnamefont{Woosley}},
  \bibinfo{author}{\bibfnamefont{R.~G.} \bibnamefont{Eastman}},
  \bibnamefont{and} \bibinfo{author}{\bibfnamefont{B.~P.}
  \bibnamefont{Schmidt}}, \bibinfo{journal}{Astrophys. J.}
  \textbf{\bibinfo{volume}{516}}, \bibinfo{pages}{788} (\bibinfo{year}{1999}).

\bibitem[{\citenamefont{Gehrels et~al.}(2005)}]{gehrels05}
\bibinfo{author}{\bibfnamefont{N.}~\bibnamefont{Gehrels}} \bibnamefont{et~al.},
  \bibinfo{journal}{Nature} \textbf{\bibinfo{volume}{437}},
  \bibinfo{pages}{851} (\bibinfo{year}{2005}).

\bibitem[{\citenamefont{Villasenor et~al.}(2005)}]{villasenor05}
\bibinfo{author}{\bibfnamefont{J.~S.} \bibnamefont{Villasenor}}
  \bibnamefont{et~al.}, \bibinfo{journal}{Nature}
  \textbf{\bibinfo{volume}{437}}, \bibinfo{pages}{855} (\bibinfo{year}{2005}).

\bibitem[{\citenamefont{Fox et~al.}(2005)}]{fox05}
\bibinfo{author}{\bibfnamefont{D.~B.} \bibnamefont{Fox}} \bibnamefont{et~al.},
  \bibinfo{journal}{Nature} \textbf{\bibinfo{volume}{437}},
  \bibinfo{pages}{845} (\bibinfo{year}{2005}).

\bibitem[{\citenamefont{Hjorth et~al.}(2005)}]{hjorth05}
\bibinfo{author}{\bibfnamefont{J.}~\bibnamefont{Hjorth}} \bibnamefont{et~al.},
  \bibinfo{journal}{Nature} \textbf{\bibinfo{volume}{437}},
  \bibinfo{pages}{859} (\bibinfo{year}{2005}).

\bibitem[{\citenamefont{Fryer et~al.}(1999)\citenamefont{Fryer, Woosley, and
  Hartmann}}]{fryer99}
\bibinfo{author}{\bibfnamefont{C.~L.} \bibnamefont{Fryer}},
  \bibinfo{author}{\bibfnamefont{S.~E.} \bibnamefont{Woosley}},
  \bibnamefont{and} \bibinfo{author}{\bibfnamefont{D.~H.}
  \bibnamefont{Hartmann}}, \bibinfo{journal}{Astrophys. J.}
  \textbf{\bibinfo{volume}{526}}, \bibinfo{pages}{152} (\bibinfo{year}{1999}).

\bibitem[{\citenamefont{Gehrels et~al.}(2006)}]{grbclass3}
\bibinfo{author}{\bibfnamefont{N.}~\bibnamefont{Gehrels}} \bibnamefont{et~al.},
  \bibinfo{journal}{Nature} \textbf{\bibinfo{volume}{444}},
  \bibinfo{pages}{1044} (\bibinfo{year}{2006}).

\bibitem[{\citenamefont{Jakobsson et~al.}(2005)}]{jakobsson04}
\bibinfo{author}{\bibfnamefont{P.}~\bibnamefont{Jakobsson}}
  \bibnamefont{et~al.}, \bibinfo{journal}{Mon. Not. R. Astron. Soc.}
  \textbf{\bibinfo{volume}{362}}, \bibinfo{pages}{245} (\bibinfo{year}{2005}).

\bibitem[{\citenamefont{Christensen et~al.}(2004)\citenamefont{Christensen,
  Hjorth, and Gorosabel}}]{christensen04}
\bibinfo{author}{\bibfnamefont{L.}~\bibnamefont{Christensen}},
  \bibinfo{author}{\bibfnamefont{J.}~\bibnamefont{Hjorth}}, \bibnamefont{and}
  \bibinfo{author}{\bibfnamefont{J.}~\bibnamefont{Gorosabel}},
  \bibinfo{journal}{Astron. \& Astrophys.} \textbf{\bibinfo{volume}{425}},
  \bibinfo{pages}{913} (\bibinfo{year}{2004}).

\bibitem[{\citenamefont{Berger et~al.}(2005)}]{berger05}
\bibinfo{author}{\bibfnamefont{E.}~\bibnamefont{Berger}} \bibnamefont{et~al.},
  \bibinfo{journal}{Astrophys. J.} \textbf{\bibinfo{volume}{634}},
  \bibinfo{pages}{501} (\bibinfo{year}{2005}).

\bibitem[{\citenamefont{Conselice et~al.}(2005)}]{conselice05}
\bibinfo{author}{\bibfnamefont{C.}~\bibnamefont{Conselice}}
  \bibnamefont{et~al.}, \bibinfo{journal}{Astrophys. J.}
  \textbf{\bibinfo{volume}{633}}, \bibinfo{pages}{29} (\bibinfo{year}{2005}).

\bibitem[{\citenamefont{Guetta and Piran}(2005)}]{guetta05}
\bibinfo{author}{\bibfnamefont{D.}~\bibnamefont{Guetta}} \bibnamefont{and}
  \bibinfo{author}{\bibfnamefont{T.}~\bibnamefont{Piran}},
  \bibinfo{journal}{Astron. \& Astrophys.} \textbf{\bibinfo{volume}{435}},
  \bibinfo{pages}{421} (\bibinfo{year}{2005}).

\bibitem[{\citenamefont{Nakar et~al.}(2006)\citenamefont{Nakar, Gal-Yam, and
  Fox}}]{NYF}
\bibinfo{author}{\bibfnamefont{E.}~\bibnamefont{Nakar}},
  \bibinfo{author}{\bibfnamefont{A.}~\bibnamefont{Gal-Yam}}, \bibnamefont{and}
  \bibinfo{author}{\bibfnamefont{D.~B.} \bibnamefont{Fox}},
  \bibinfo{journal}{Astrophys. J.} \textbf{\bibinfo{volume}{650}},
  \bibinfo{pages}{281} (\bibinfo{year}{2006}).

\bibitem[{\citenamefont{Guetta and Piran}(2006)}]{guetta06}
\bibinfo{author}{\bibfnamefont{D.}~\bibnamefont{Guetta}} \bibnamefont{and}
  \bibinfo{author}{\bibfnamefont{T.}~\bibnamefont{Piran}},
  \bibinfo{journal}{Astron. \& Astrophys.} \textbf{\bibinfo{volume}{453}},
  \bibinfo{pages}{823} (\bibinfo{year}{2006}).

\bibitem[{\citenamefont{Ott et~al.}(2004)\citenamefont{Ott, Burrows, Livne, and
  Walder}}]{ottburrows04}
\bibinfo{author}{\bibfnamefont{C.}~\bibnamefont{Ott}},
  \bibinfo{author}{\bibfnamefont{A.}~\bibnamefont{Burrows}},
  \bibinfo{author}{\bibfnamefont{E.}~\bibnamefont{Livne}}, \bibnamefont{and}
  \bibinfo{author}{\bibfnamefont{R.}~\bibnamefont{Walder}},
  \bibinfo{journal}{Astrophys. J.} \textbf{\bibinfo{volume}{600}},
  \bibinfo{pages}{834} (\bibinfo{year}{2004}).

\bibitem[{\citenamefont{Dimmelmeier et~al.}(2002)\citenamefont{Dimmelmeier,
  Font, and M{\"{u}}ller}}]{dfm02}
\bibinfo{author}{\bibfnamefont{H.}~\bibnamefont{Dimmelmeier}},
  \bibinfo{author}{\bibfnamefont{J.~A.} \bibnamefont{Font}}, \bibnamefont{and}
  \bibinfo{author}{\bibfnamefont{E.}~\bibnamefont{M{\"{u}}ller}},
  \bibinfo{journal}{Astron. \& Astrophys.} \textbf{\bibinfo{volume}{393}},
  \bibinfo{pages}{523} (\bibinfo{year}{2002}).

\bibitem[{\citenamefont{Flanagan and Hughes}(1998)}]{flanagan98}
\bibinfo{author}{\bibfnamefont{{\`{E}}.~{\`{E}}.} \bibnamefont{Flanagan}}
  \bibnamefont{and} \bibinfo{author}{\bibfnamefont{S.~A.}
  \bibnamefont{Hughes}}, \bibinfo{journal}{Phys. Rev. D}
  \textbf{\bibinfo{volume}{57}}, \bibinfo{pages}{4535} (\bibinfo{year}{1998}).

\bibitem[{\citenamefont{Sperhake et~al.}(2005)\citenamefont{Sperhake, Kelly,
  Laguna, Smith, and Schnetter}}]{BHBHmerge1}
\bibinfo{author}{\bibfnamefont{U.}~\bibnamefont{Sperhake}},
  \bibinfo{author}{\bibfnamefont{B.}~\bibnamefont{Kelly}},
  \bibinfo{author}{\bibfnamefont{P.}~\bibnamefont{Laguna}},
  \bibinfo{author}{\bibfnamefont{K.~L.} \bibnamefont{Smith}}, \bibnamefont{and}
  \bibinfo{author}{\bibfnamefont{E.}~\bibnamefont{Schnetter}},
  \bibinfo{journal}{\prd} \textbf{\bibinfo{volume}{71}},
  \bibinfo{pages}{124042} (\bibinfo{year}{2005}).

\bibitem[{\citenamefont{Baker et~al.}(2006)\citenamefont{Baker, Centrella,
  Choi, Koppitz, and van Meter}}]{BHBHmerge2}
\bibinfo{author}{\bibfnamefont{J.~G.} \bibnamefont{Baker}},
  \bibinfo{author}{\bibfnamefont{J.}~\bibnamefont{Centrella}},
  \bibinfo{author}{\bibfnamefont{D.~I.} \bibnamefont{Choi}},
  \bibinfo{author}{\bibfnamefont{M.}~\bibnamefont{Koppitz}}, \bibnamefont{and}
  \bibinfo{author}{\bibfnamefont{J.}~\bibnamefont{van Meter}},
  \bibinfo{journal}{\prd} \textbf{\bibinfo{volume}{73}},
  \bibinfo{pages}{104002} (\bibinfo{year}{2006}).

\bibitem[{\citenamefont{Cadonati}(2004)}]{cadonati04}
\bibinfo{author}{\bibfnamefont{L.}~\bibnamefont{Cadonati}},
  \bibinfo{journal}{Classical Quantum Gravity} \textbf{\bibinfo{volume}{21}},
  \bibinfo{pages}{S1695} (\bibinfo{year}{2004}).

\bibitem[{\citenamefont{Abbott et~al.}(2005{\natexlab{a}})}]{abbottgrb05}
\bibinfo{author}{\bibfnamefont{B.}~\bibnamefont{Abbott}} \bibnamefont{et~al.},
  \bibinfo{journal}{Phys. Rev. D} \textbf{\bibinfo{volume}{72}},
  \bibinfo{pages}{042002} (\bibinfo{year}{2005}{\natexlab{a}}).

\bibitem[{\citenamefont{Abbott et~al.}(2004)}]{abbottnim04}
\bibinfo{author}{\bibfnamefont{B.}~\bibnamefont{Abbott}} \bibnamefont{et~al.},
  \bibinfo{journal}{Nucl. Inst. \& Meth. in Phys. Res.}
  \textbf{\bibinfo{volume}{517}}, \bibinfo{pages}{154} (\bibinfo{year}{2004}).

\bibitem[{\citenamefont{Hurley and Murdin}(2002)}]{ipn02}
\bibinfo{author}{\bibfnamefont{K.}~\bibnamefont{Hurley}} \bibnamefont{and}
  \bibinfo{author}{\bibfnamefont{P.}~\bibnamefont{Murdin}},
  \bibinfo{journal}{Encyclopedia of Astronomy and Astrophysics}
  (\bibinfo{year}{2002}).

\bibitem[{\citenamefont{Aptekar et~al.}(1995)}]{konus95}
\bibinfo{author}{\bibfnamefont{R.}~\bibnamefont{Aptekar}} \bibnamefont{et~al.},
  \bibinfo{journal}{Space Science Reviews} \textbf{\bibinfo{volume}{71}},
  \bibinfo{pages}{265} (\bibinfo{year}{1995}).

\bibitem[{\citenamefont{Rees and M{\'{e}}sz{\'{a}}ros}(1994)}]{rees94}
\bibinfo{author}{\bibfnamefont{M.~J.} \bibnamefont{Rees}} \bibnamefont{and}
  \bibinfo{author}{\bibfnamefont{P.}~\bibnamefont{M{\'{e}}sz{\'{a}}ros}},
  \bibinfo{journal}{Astrophys. J.} \textbf{\bibinfo{volume}{430}},
  \bibinfo{pages}{L93} (\bibinfo{year}{1994}).

\bibitem[{\citenamefont{Kochanek and Piran}(1993)}]{kochanek93}
\bibinfo{author}{\bibfnamefont{C.~S.} \bibnamefont{Kochanek}} \bibnamefont{and}
  \bibinfo{author}{\bibfnamefont{T.}~\bibnamefont{Piran}},
  \bibinfo{journal}{Astrophys. J.} \textbf{\bibinfo{volume}{417}},
  \bibinfo{pages}{L17} (\bibinfo{year}{1993}).

\bibitem[{\citenamefont{Piran}(1999)}]{piran99}
\bibinfo{author}{\bibfnamefont{T.}~\bibnamefont{Piran}},
  \bibinfo{journal}{Phys. Rep.} \textbf{\bibinfo{volume}{314}},
  \bibinfo{pages}{575} (\bibinfo{year}{1999}).

\bibitem[{\citenamefont{M{\'{e}}sz{\'{a}}ros}(2006)}]{meszaros06}
\bibinfo{author}{\bibfnamefont{P.}~\bibnamefont{M{\'{e}}sz{\'{a}}ros}},
  \bibinfo{journal}{Rep. Prog. Phys.} \textbf{\bibinfo{volume}{69}},
  \bibinfo{pages}{2259} (\bibinfo{year}{2006}).

\bibitem[{\citenamefont{Finn et~al.}(1999)\citenamefont{Finn, Mohanty, and
  Romano}}]{FMR}
\bibinfo{author}{\bibfnamefont{L.~S.} \bibnamefont{Finn}},
  \bibinfo{author}{\bibfnamefont{S.~D.} \bibnamefont{Mohanty}},
  \bibnamefont{and} \bibinfo{author}{\bibfnamefont{J.~D.}
  \bibnamefont{Romano}}, \bibinfo{journal}{\prd} \textbf{\bibinfo{volume}{60}},
  \bibinfo{pages}{121101(R)} (\bibinfo{year}{1999}).

\bibitem[{\citenamefont{Astone et~al.}(2002)}]{astone1}
\bibinfo{author}{\bibfnamefont{P.}~\bibnamefont{Astone}} \bibnamefont{et~al.},
  \bibinfo{journal}{\prd} \textbf{\bibinfo{volume}{66}},
  \bibinfo{pages}{102002} (\bibinfo{year}{2002}).

\bibitem[{\citenamefont{Astone et~al.}(2005)}]{astone2}
\bibinfo{author}{\bibfnamefont{P.}~\bibnamefont{Astone}} \bibnamefont{et~al.},
  \bibinfo{journal}{\prd} \textbf{\bibinfo{volume}{71}},
  \bibinfo{pages}{042001} (\bibinfo{year}{2005}).

\bibitem[{\citenamefont{Helstrom}(1968)}]{helstrom}
\bibinfo{author}{\bibfnamefont{C.~W.} \bibnamefont{Helstrom}},
  \emph{\bibinfo{title}{Statistical Theory of Signal Detection}}
  (\bibinfo{publisher}{Pergamon Press, London}, \bibinfo{year}{1968}),
  \bibinfo{edition}{2nd} ed.

\bibitem[{\citenamefont{Mohanty}(2005)}]{SDM:gwdaw9}
\bibinfo{author}{\bibfnamefont{S.}~\bibnamefont{Mohanty}},
  \bibinfo{journal}{Classical Quantum Gravity} \textbf{\bibinfo{volume}{22}},
  \bibinfo{pages}{S1349} (\bibinfo{year}{2005}).

\bibitem[{\citenamefont{Lehman}(1998)}]{ranksum}
\bibinfo{author}{\bibfnamefont{E.~L.} \bibnamefont{Lehman}},
  \emph{\bibinfo{title}{Nonparametrics}} (\bibinfo{publisher}{Prentice Hall},
  \bibinfo{year}{1998}).

\bibitem[{\citenamefont{Thorne}(1987)}]{300years}
\bibinfo{author}{\bibfnamefont{K.~S.} \bibnamefont{Thorne}}, in
  \emph{\bibinfo{booktitle}{300 Years of Gravitation}} (\bibinfo{year}{1987}),
  pp. \bibinfo{pages}{330--457}.

\bibitem[{\citenamefont{{Particle Data Group}}(2002)}]{pdg02}
\bibinfo{author}{\bibnamefont{{Particle Data Group}}}, \bibinfo{journal}{\prd}
  \textbf{\bibinfo{volume}{66}}, \bibinfo{pages}{010001}
  (\bibinfo{year}{2002}).

\bibitem[{\citenamefont{Mohanty}(2006)}]{SDM:GWDAW10}
\bibinfo{author}{\bibfnamefont{S.~D.} \bibnamefont{Mohanty}},
  \bibinfo{journal}{Classical Quantum Gravity} \textbf{\bibinfo{volume}{23}},
  \bibinfo{pages}{S723} (\bibinfo{year}{2006}).

\bibitem[{\citenamefont{Bromm and Loeb}(2002)}]{bromm+loeb:2002}
\bibinfo{author}{\bibfnamefont{V.}~\bibnamefont{Bromm}} \bibnamefont{and}
  \bibinfo{author}{\bibfnamefont{A.}~\bibnamefont{Loeb}},
  \bibinfo{journal}{Astrophys. J.} \textbf{\bibinfo{volume}{575}},
  \bibinfo{pages}{111} (\bibinfo{year}{2002}).

\bibitem[{\citenamefont{Bagolyi et~al.}(2006)}]{bagolyi06}
\bibinfo{author}{\bibfnamefont{Z.}~\bibnamefont{Bagolyi}} \bibnamefont{et~al.},
  \bibinfo{journal}{Astron. \& Astrophys.} \textbf{\bibinfo{volume}{453}},
  \bibinfo{pages}{797} (\bibinfo{year}{2006}).

\bibitem[{\citenamefont{Isaacson}(1968)}]{isaacson}
\bibinfo{author}{\bibfnamefont{R.~A.} \bibnamefont{Isaacson}},
  \bibinfo{journal}{Phys. Rev.} \textbf{\bibinfo{volume}{166}},
  \bibinfo{pages}{1272} (\bibinfo{year}{1968}).

\bibitem[{\citenamefont{Fruchter et~al.}(2006)}]{LGRBvsSN}
\bibinfo{author}{\bibfnamefont{A.~S.} \bibnamefont{Fruchter}}
  \bibnamefont{et~al.}, \bibinfo{journal}{Nature}
  \textbf{\bibinfo{volume}{441}}, \bibinfo{pages}{463} (\bibinfo{year}{2006}).

\bibitem[{\citenamefont{Fryer et~al.}(2002)\citenamefont{Fryer, Holz, and
  Hughes}}]{Fryer}
\bibinfo{author}{\bibfnamefont{C.~L.} \bibnamefont{Fryer}},
  \bibinfo{author}{\bibfnamefont{D.~E.} \bibnamefont{Holz}}, \bibnamefont{and}
  \bibinfo{author}{\bibfnamefont{S.~A.} \bibnamefont{Hughes}},
  \bibinfo{journal}{Astrophys. J.} \textbf{\bibinfo{volume}{565}},
  \bibinfo{pages}{430} (\bibinfo{year}{2002}).

\bibitem[{\citenamefont{Dyson}(1963)}]{dyson}
\bibinfo{author}{\bibfnamefont{F.}~\bibnamefont{Dyson}}, in
  \emph{\bibinfo{booktitle}{Interstellar Communication}}, edited by
  \bibinfo{editor}{\bibfnamefont{A.~G.~W.} \bibnamefont{Cameron}}
  (\bibinfo{year}{1963}).

\bibitem[{\citenamefont{Shibata et~al.}(2006)\citenamefont{Shibata, Duez, Liu,
  Shapiro, and Stephens}}]{HMNS}
\bibinfo{author}{\bibfnamefont{M.}~\bibnamefont{Shibata}},
  \bibinfo{author}{\bibfnamefont{M.~D.} \bibnamefont{Duez}},
  \bibinfo{author}{\bibfnamefont{Y.~T.} \bibnamefont{Liu}},
  \bibinfo{author}{\bibfnamefont{S.~L.} \bibnamefont{Shapiro}},
  \bibnamefont{and} \bibinfo{author}{\bibfnamefont{B.~C.}
  \bibnamefont{Stephens}}, \bibinfo{journal}{\prl}
  \textbf{\bibinfo{volume}{96}}, \bibinfo{pages}{031102}
  (\bibinfo{year}{2006}).

\bibitem[{\citenamefont{Shibata and Taniguchi}(2006)}]{NSNSmerge2}
\bibinfo{author}{\bibfnamefont{M.}~\bibnamefont{Shibata}} \bibnamefont{and}
  \bibinfo{author}{\bibfnamefont{K.}~\bibnamefont{Taniguchi}},
  \bibinfo{journal}{\prd} \textbf{\bibinfo{volume}{73}},
  \bibinfo{pages}{064027} (\bibinfo{year}{2006}).

\bibitem[{\citenamefont{Abbott et~al.}(2005{\natexlab{b}})}]{LIGOinspiral}
\bibinfo{author}{\bibfnamefont{B.}~\bibnamefont{Abbott}} \bibnamefont{et~al.},
  \bibinfo{journal}{\prd} \textbf{\bibinfo{volume}{72}},
  \bibinfo{pages}{082001} (\bibinfo{year}{2005}{\natexlab{b}}).

\bibitem[{\citenamefont{Shibata and Uryu}(2002)}]{NSNSmerge}
\bibinfo{author}{\bibfnamefont{M.}~\bibnamefont{Shibata}} \bibnamefont{and}
  \bibinfo{author}{\bibfnamefont{K.}~\bibnamefont{Uryu}},
  \bibinfo{journal}{Prog. Theor. Phys.} \textbf{\bibinfo{volume}{107}},
  \bibinfo{pages}{265} (\bibinfo{year}{2002}).

\bibitem[{\citenamefont{Oechslin and Janka}(2007)}]{oechslin07}
\bibinfo{author}{\bibfnamefont{R.}~\bibnamefont{Oechslin}} \bibnamefont{and}
  \bibinfo{author}{\bibfnamefont{H.-T.} \bibnamefont{Janka}},
  \bibinfo{journal}{arXiv:astro-ph/0702228}  (\bibinfo{year}{2007}).

\bibitem[{\citenamefont{Loffler et~al.}(2006)\citenamefont{Loffler, Rezzolla,
  and Ansorg}}]{BHNSmerge1}
\bibinfo{author}{\bibfnamefont{F.}~\bibnamefont{Loffler}},
  \bibinfo{author}{\bibfnamefont{L.}~\bibnamefont{Rezzolla}}, \bibnamefont{and}
  \bibinfo{author}{\bibfnamefont{M.}~\bibnamefont{Ansorg}},
  \bibinfo{journal}{\prd} \textbf{\bibinfo{volume}{74}},
  \bibinfo{pages}{104018} (\bibinfo{year}{2006}).

\bibitem[{\citenamefont{Shibata and Uryu}(2006)}]{BHNSmerge2}
\bibinfo{author}{\bibfnamefont{M.}~\bibnamefont{Shibata}} \bibnamefont{and}
  \bibinfo{author}{\bibfnamefont{K.}~\bibnamefont{Uryu}},
  \bibinfo{journal}{\prd} \textbf{\bibinfo{volume}{74}},
  \bibinfo{pages}{121503(R)} (\bibinfo{year}{2006}).

\bibitem[{\citenamefont{Faber et~al.}(2006)}]{faber06}
\bibinfo{author}{\bibfnamefont{J.~A.} \bibnamefont{Faber}}
  \bibnamefont{et~al.}, \bibinfo{journal}{Astrophys. J.}
  \textbf{\bibinfo{volume}{641}}, \bibinfo{pages}{L93} (\bibinfo{year}{2006}).

\bibitem[{\citenamefont{Klimenko et~al.}(2005)\citenamefont{Klimenko, Mohanty,
  Rakhmanov, and Mitselmakher}}]{klimenko+etal:05}
\bibinfo{author}{\bibfnamefont{S.}~\bibnamefont{Klimenko}},
  \bibinfo{author}{\bibfnamefont{S.}~\bibnamefont{Mohanty}},
  \bibinfo{author}{\bibfnamefont{M.}~\bibnamefont{Rakhmanov}},
  \bibnamefont{and}
  \bibinfo{author}{\bibfnamefont{G.}~\bibnamefont{Mitselmakher}},
  \bibinfo{journal}{\prd} \textbf{\bibinfo{volume}{72}},
  \bibinfo{pages}{122002} (\bibinfo{year}{2005}).

\bibitem[{\citenamefont{Rakhmanov}(2006)}]{Rakhmanov:gwdaw10}
\bibinfo{author}{\bibfnamefont{M.}~\bibnamefont{Rakhmanov}},
  \bibinfo{journal}{Classical Quantum Gravity} \textbf{\bibinfo{volume}{23}},
  \bibinfo{pages}{S673} (\bibinfo{year}{2006}).

\bibitem[{\citenamefont{Mohanty et~al.}(2006)}]{mohanty+etal:06}
\bibinfo{author}{\bibfnamefont{S.~D.} \bibnamefont{Mohanty}}
  \bibnamefont{et~al.}, \bibinfo{journal}{Classical Quantum Gravity}
  \textbf{\bibinfo{volume}{23}}, \bibinfo{pages}{4799} (\bibinfo{year}{2006}).

\bibitem[{\citenamefont{van Putten et~al.}(2004)}]{vanPutten:grb}
\bibinfo{author}{\bibfnamefont{M.~H.} \bibnamefont{van Putten}}
  \bibnamefont{et~al.}, \bibinfo{journal}{\prd} \textbf{\bibinfo{volume}{69}},
  \bibinfo{pages}{044007} (\bibinfo{year}{2004}).

\end{thebibliography}


\clearpage
\begin{turnpage}
\begin{table}
\newcolumntype{A}{>{\raggedright\arraybackslash}m{1.4cm}}
\newcolumntype{f}[1]{D{.}{.}{#1}}
\caption{\label{tab:s4ul25ms}S4 90\% upper limits on hrss of Q = 8.9 linearly polarized sine-gaussians, in units of
$10^{-21}\:\text{Hz}^{-1/2}$; 25-ms crosscorrelation length.}
\begin{tabular}{|c|c|c|c|c|c|c|c|c|c|c|c|c|c|c|c|c|c|c|c|}
\hline
\hline
                           & \multicolumn{3}{c|}{100 Hz} & \multicolumn{3}{c|}{150 Hz}  & \multicolumn{3}{c|}{250 Hz}  &
                             \multicolumn{3}{c|}{554 Hz} & \multicolumn{3}{c|}{1000 Hz} & \multicolumn{3}{c|}{1850 Hz} \\\cline{2-19}
GRB     &      &      &      &      &      &      &      &      &      &      &      &      &      &      &      &      &      &      \\
date & H1-H2 & H1-L1 & H2-L1 & H1-H2 & H1-L1 & H2-L1 &
       H1-H2 & H1-L1 & H2-L1 & H1-H2 & H1-L1 & H2-L1 &
       H1-H2 & H1-L1 & H2-L1 & H1-H2 & H1-L1 & H2-L1 \\
\hline
\hline
050223  &  5.5 &  ... &  ... &  3.6 &  ... &  ... &  4.1 &  ... &  ... &  6.9 &  ... &  ... & 11.7 &  ... &  ... & 25.8 &  ... &  ... \\
050306  &  7.8 &  6.4 & 12.0 &  5.2 &  5.2 &  8.8 &  5.6 &  6.3 &  9.5 &  9.0 & 12.6 & 16.0 & 16.4 & 24.5 & 30.4 & 31.4 & 61.9 & 82.4 \\
050318  &  7.9 & 10.2 & 15.4 &  6.0 &  7.0 & 10.7 &  6.0 &  9.3 & 11.9 &  9.5 & 16.7 & 19.8 & 15.8 & 30.2 & 35.0 & 33.4 & 55.3 & 66.7 \\
050319  &  6.6 &  6.8 &  8.3 &  4.7 &  4.9 &  5.7 &  5.4 &  6.1 &  6.2 &  8.1 & 11.1 & 11.0 & 15.5 & 21.1 & 19.8 & 29.7 & 36.9 & 34.9 \\
\hline
\end{tabular}

\caption{\label{tab:s3ul25ms}S3 90\% upper limits on hrss of Q = 8.9 linearly polarized sine-gaussians, in units of
$10^{-20}\:\text{Hz}^{-1/2}$; 25-ms crosscorrelation length.}
\begin{tabular}{|c|c|c|c|c|c|c|c|c|c|c|c|c|c|c|c|c|c|c|c|}
\hline
\hline
                           & \multicolumn{3}{c|}{100 Hz} & \multicolumn{3}{c|}{150 Hz}  & \multicolumn{3}{c|}{250 Hz}  &
                             \multicolumn{3}{c|}{554 Hz} & \multicolumn{3}{c|}{1000 Hz} & \multicolumn{3}{c|}{1850 Hz} \\\cline{2-19}
GRB     &      &      &      &      &      &      &      &      &      &      &      &      &      &      &      &      &      &      \\
date & H1-H2 & H1-L1 & H2-L1 & H1-H2 & H1-L1 & H2-L1 &
       H1-H2 & H1-L1 & H2-L1 & H1-H2 & H1-L1 & H2-L1 &
       H1-H2 & H1-L1 & H2-L1 & H1-H2 & H1-L1 & H2-L1 \\
\hline
\hline
031108  &  6.5 &  ... &  ... &  3.6 &  ... &  ... &  3.6 &  ... &  ... &  4.2 &  ... &  ... &  6.7 &  ... &  ... & 19.7 &  ... &  ... \\
031109a &  4.8 &  ... &  ... &  2.9 &  ... &  ... &  2.9 &  ... &  ... &  3.6 &  ... &  ... &  6.0 &  ... &  ... & 14.7 &  ... &  ... \\
031220  &  5.7 &  ... &  ... &  3.3 &  ... &  ... &  3.0 &  ... &  ... &  3.7 &  ... &  ... &  6.3 &  ... &  ... & 14.7 &  ... &  ... \\
\hline
\end{tabular}

\caption{\label{tab:s2ul25ms}S2 90\% upper limits on hrss of Q = 8.9 linearly polarized sine-gaussians, in units of
$10^{-19}\:\text{Hz}^{-1/2}$; 25-ms crosscorrelation length.}
\begin{tabular}{|c|c|c|c|c|c|c|c|c|c|c|c|c|c|c|c|c|c|c|c|}
\hline
\hline
                           & \multicolumn{3}{c|}{100 Hz} & \multicolumn{3}{c|}{150 Hz}  & \multicolumn{3}{c|}{250 Hz}  &
                             \multicolumn{3}{c|}{554 Hz} & \multicolumn{3}{c|}{1000 Hz} & \multicolumn{3}{c|}{1850 Hz} \\\cline{2-19}
GRB     &      &      &      &      &      &      &      &      &      &      &      &      &      &      &      &      &      &      \\
date & H1-H2 & H1-L1 & H2-L1 & H1-H2 & H1-L1 & H2-L1 &
       H1-H2 & H1-L1 & H2-L1 & H1-H2 & H1-L1 & H2-L1 &
       H1-H2 & H1-L1 & H2-L1 & H1-H2 & H1-L1 & H2-L1 \\
\hline
\hline
030217  &  ... &  ... &  4.4 &  ... &  ... &  2.2 &  ... &  ... &  1.0 &  ... &  ... &  1.6 &  ... &  ... &  4.4 &  ... &  ... & 10.2 \\
030226  &  7.7 &  3.5 &  5.4 &  3.4 &  1.6 &  2.2 & 1.00 & 0.68 & 0.63 &  1.3 &  1.1 & 0.81 &  2.6 &  2.4 &  1.4 &  7.1 &  6.6 &  2.7 \\
030320a &  7.2 &  2.1 &  7.1 &  2.5 &  1.1 &  2.2 & 0.69 & 0.58 & 0.71 &  1.0 &  1.1 &  1.3 &  1.6 &  2.9 &  3.1 &  3.8 &  6.0 &  5.6 \\
030323a &  5.1 &  3.1 &  6.4 &  2.5 &  1.7 &  2.9 &  1.1 & 0.99 &  1.5 &  1.7 &  2.3 &  3.3 &  2.6 &  6.1 &  7.2 &  6.0 & 11.4 & 13.4 \\
030323b &  4.6 &  1.8 &  5.2 &  1.7 & 0.94 &  1.8 & 0.64 & 0.45 & 0.81 & 0.92 & 0.82 &  1.5 &  1.3 &  1.8 &  2.4 &  3.0 &  3.5 &  4.8 \\
030324  &  9.2 &  ... &  ... &  4.7 &  ... &  ... &  1.6 &  ... &  ... &  2.0 &  ... &  ... &  3.3 &  ... &  ... &  7.9 &  ... &  ... \\
030325  &  2.8 &  1.7 &  3.0 &  1.3 & 0.80 &  1.5 & 0.55 & 0.48 & 0.76 & 0.89 &  1.0 &  1.5 &  1.3 &  2.0 &  2.4 &  3.2 &  4.9 &  5.3 \\
030326  & 10.2 &  3.9 &  9.6 &  4.4 &  2.1 &  3.7 &  1.4 & 0.94 &  1.2 &  2.0 &  1.6 &  1.9 &  3.1 &  3.4 &  3.1 &  8.4 &  8.1 &  6.3 \\
030329a &  4.6 &  ... &  ... &  2.4 &  ... &  ... &  1.1 &  ... &  ... &  1.8 &  ... &  ... &  3.0 &  ... &  ... &  7.6 &  ... &  ... \\
030329b &  2.8 &  ... &  ... &  1.1 &  ... &  ... & 0.31 &  ... &  ... & 0.55 &  ... &  ... & 0.89 &  ... &  ... &  2.0 &  ... &  ... \\
030331  &  ... &  3.4 &  ... &  ... &  1.6 &  ... &  ... & 0.85 &  ... &  ... &  2.0 &  ... &  ... &  3.4 &  ... &  ... &  8.0 &  ... \\
030405  &  2.1 &  1.4 &  3.1 &  1.0 & 0.80 &  1.3 & 0.34 & 0.42 & 0.51 & 0.59 & 0.76 & 0.97 & 0.87 &  2.0 &  2.2 &  2.0 &  4.8 &  4.5 \\
030406  &  ... &  1.2 &  ... &  ... & 0.67 &  ... &  ... & 0.42 &  ... &  ... & 0.77 &  ... &  ... &  1.7 &  ... &  ... &  4.4 &  ... \\
030413  &  ... &  ... &  1.6 &  ... &  ... & 0.85 &  ... &  ... & 0.50 &  ... &  ... & 0.89 &  ... &  ... &  2.3 &  ... &  ... &  4.4 \\
030414  &  1.4 &  ... &  ... & 0.91 &  ... &  ... & 0.32 &  ... &  ... & 0.39 &  ... &  ... & 0.70 &  ... &  ... &  1.6 &  ... &  ... \\
\hline
\end{tabular}
\end{table}
\end{turnpage}

\begin{turnpage}
\begin{table}
\caption{\label{tab:s4ul25mspsic}S4 90\% upper limits on hrss of Q = 8.9 circularly polarized sine-gaussians, in units of
$10^{-21}\:\text{Hz}^{-1/2}$; 25-ms crosscorrelation length.}
\begin{tabular}{|c|c|c|c|c|c|c|c|c|c|c|c|c|c|c|c|c|c|c|c|}
\hline
\hline
                           & \multicolumn{3}{c|}{100 Hz} & \multicolumn{3}{c|}{150 Hz}  & \multicolumn{3}{c|}{250 Hz}  &
                             \multicolumn{3}{c|}{554 Hz} & \multicolumn{3}{c|}{1000 Hz} & \multicolumn{3}{c|}{1850 Hz} \\\cline{2-19}
GRB     &      &      &      &      &      &      &      &      &      &      &      &      &      &      &      &      &      &      \\
date & H1-H2 & H1-L1 & H2-L1 & H1-H2 & H1-L1 & H2-L1 &
       H1-H2 & H1-L1 & H2-L1 & H1-H2 & H1-L1 & H2-L1 &
       H1-H2 & H1-L1 & H2-L1 & H1-H2 & H1-L1 & H2-L1 \\
\hline
\hline
050223  &  1.6 &  ... &  ... &  1.1 &  ... &  ... &  1.2 &  ... &  ... &  2.0 &  ... &  ... &  3.5 &  ... &  ... &  6.7 &  ... &  ... \\
050306  &  2.2 &  1.4 &  2.6 &  1.5 &  1.1 &  1.8 &  1.6 &  1.4 &  2.0 &  2.6 &  2.6 &  3.3 &  4.5 &  5.0 &  6.2 &  8.5 & 14.2 & 17.6 \\
050318  &  2.2 &  2.2 &  3.1 &  1.6 &  1.5 &  2.2 &  1.6 &  1.9 &  2.4 &  2.6 &  3.5 &  4.0 &  4.6 &  6.1 &  6.9 &  8.8 & 11.1 & 13.1 \\
050319  &  1.8 &  1.8 &  2.3 &  1.4 &  1.3 &  1.6 &  1.5 &  1.7 &  1.8 &  2.4 &  3.1 &  3.0 &  4.3 &  5.5 &  5.2 &  8.2 & 10.0 &  9.9 \\
\hline
\end{tabular}

\caption{\label{tab:s3ul25mspsic}S3 90\% upper limits on hrss of Q = 8.9 circularly polarized sine-gaussians, in units of
$10^{-21}\:\text{Hz}^{-1/2}$; 25-ms crosscorrelation length.}
\begin{tabular}{|c|c|c|c|c|c|c|c|c|c|c|c|c|c|c|c|c|c|c|c|}
\hline
\hline
                           & \multicolumn{3}{c|}{100 Hz} & \multicolumn{3}{c|}{150 Hz}  & \multicolumn{3}{c|}{250 Hz}  &
                             \multicolumn{3}{c|}{554 Hz} & \multicolumn{3}{c|}{1000 Hz} & \multicolumn{3}{c|}{1850 Hz} \\\cline{2-19}
GRB     &      &      &      &      &      &      &      &      &      &      &      &      &      &      &      &      &      &      \\
date & H1-H2 & H1-L1 & H2-L1 & H1-H2 & H1-L1 & H2-L1 &
       H1-H2 & H1-L1 & H2-L1 & H1-H2 & H1-L1 & H2-L1 &
       H1-H2 & H1-L1 & H2-L1 & H1-H2 & H1-L1 & H2-L1 \\
\hline
\hline
031108  & 19.0 &  ... &  ... & 11.3 &  ... &  ... & 10.9 &  ... &  ... & 12.5 &  ... &  ... & 20.4 &  ... &  ... & 51.5 &  ... &  ... \\
031109a & 14.7 &  ... &  ... &  8.8 &  ... &  ... &  8.5 &  ... &  ... & 10.6 &  ... &  ... & 17.3 &  ... &  ... & 42.2 &  ... &  ... \\
031220  & 14.4 &  ... &  ... & 10.1 &  ... &  ... &  8.9 &  ... &  ... & 10.8 &  ... &  ... & 18.4 &  ... &  ... & 42.7 &  ... &  ... \\
\hline
\end{tabular}

\caption{\label{tab:s2ul25mspsic}S2 90\% upper limits on hrss of Q = 8.9 circularly polarized sine-gaussians, in units of
$10^{-20}\:\text{Hz}^{-1/2}$; 25-ms crosscorrelation length.}
\begin{tabular}{|c|c|c|c|c|c|c|c|c|c|c|c|c|c|c|c|c|c|c|c|}
\hline
\hline
                           & \multicolumn{3}{c|}{100 Hz} & \multicolumn{3}{c|}{150 Hz}  & \multicolumn{3}{c|}{250 Hz}  &
                             \multicolumn{3}{c|}{554 Hz} & \multicolumn{3}{c|}{1000 Hz} & \multicolumn{3}{c|}{1850 Hz} \\\cline{2-19}
GRB     &      &      &      &      &      &      &      &      &      &      &      &      &      &      &      &      &      &      \\
date & H1-H2 & H1-L1 & H2-L1 & H1-H2 & H1-L1 & H2-L1 &
       H1-H2 & H1-L1 & H2-L1 & H1-H2 & H1-L1 & H2-L1 &
       H1-H2 & H1-L1 & H2-L1 & H1-H2 & H1-L1 & H2-L1 \\
\hline
\hline
030226  & 22.2 & 11.0 & 18.0 &  9.2 &  5.0 &  6.9 &  2.9 &  2.1 &  1.9 &  3.7 &  3.3 &  2.6 &  7.1 &  7.1 &  4.1 & 20.3 & 20.3 &  7.2 \\
030320a & 21.9 &  7.0 & 26.6 &  7.3 &  3.6 &  7.9 &  2.0 &  1.9 &  2.2 &  2.9 &  3.3 &  4.1 &  4.6 &  9.5 & 10.1 & 10.7 & 17.3 & 16.1 \\
030323a & 16.1 &  ... &  ... &  7.9 &  ... &  ... &  3.6 &  ... &  ... &  5.6 &  ... &  ... &  7.9 &  ... &  ... & 18.5 &  ... &  ... \\
030323b & 13.4 &  4.9 & 15.5 &  4.9 &  2.5 &  5.1 &  1.8 &  1.2 &  2.3 &  2.7 &  2.4 &  4.0 &  3.7 &  5.1 &  6.6 &  8.5 &  9.2 & 12.3 \\
030324  & 28.0 &  ... &  ... & 13.3 &  ... &  ... &  4.3 &  ... &  ... &  5.6 &  ... &  ... &  9.4 &  ... &  ... & 22.2 &  ... &  ... \\
030325  &  9.0 &  4.3 &  9.5 &  4.0 &  2.0 &  4.2 &  2.0 &  1.2 &  2.4 &  3.1 &  2.8 &  4.4 &  4.3 &  5.3 &  6.7 & 10.2 & 12.2 & 15.0 \\
030326  & 29.7 & 15.1 & 39.9 & 12.4 &  8.1 & 14.9 &  4.0 &  3.5 &  4.8 &  5.8 &  5.8 &  7.6 &  9.6 & 12.1 & 11.7 & 24.2 & 25.8 & 19.7 \\
030329a & 13.8 &  ... &  ... &  7.3 &  ... &  ... &  3.3 &  ... &  ... &  5.1 &  ... &  ... &  8.2 &  ... &  ... & 21.6 &  ... &  ... \\
030329b &  8.8 &  ... &  ... &  3.2 &  ... &  ... & 0.90 &  ... &  ... &  1.5 &  ... &  ... &  2.4 &  ... &  ... &  5.9 &  ... &  ... \\
030331  &  ... &  7.1 &  ... &  ... &  3.5 &  ... &  ... &  1.8 &  ... &  ... &  4.3 &  ... &  ... &  7.3 &  ... &  ... & 17.4 &  ... \\
030405  &  6.2 &  3.4 &  8.2 &  2.9 &  2.0 &  3.4 & 0.99 &  1.1 &  1.3 &  1.6 &  2.0 &  2.5 &  2.5 &  5.1 &  5.4 &  5.9 & 11.3 & 10.7 \\
030406  &  ... &  2.8 &  ... &  ... &  1.5 &  ... &  ... & 0.90 &  ... &  ... &  1.8 &  ... &  ... &  4.0 &  ... &  ... & 10.0 &  ... \\
030413  &  ... &  ... &  4.1 &  ... &  ... &  2.2 &  ... &  ... &  1.3 &  ... &  ... &  2.4 &  ... &  ... &  6.0 &  ... &  ... & 11.0 \\
030414  &  4.1 &  ... &  ... &  2.6 &  ... &  ... & 0.82 &  ... &  ... &  1.1 &  ... &  ... &  1.9 &  ... &  ... &  4.6 &  ... &  ... \\
\hline
\end{tabular}
\end{table}
\end{turnpage}


\begin{turnpage}
\begin{table}
\newcolumntype{A}{>{\raggedright\arraybackslash}m{1.4cm}}
\newcolumntype{f}[1]{D{.}{.}{#1}}
\caption{\label{tab:s4ul100ms}S4 90\% upper limits on hrss of Q = 8.9 linearly polarized sine-gaussians, in units of
$10^{-21}\:\text{Hz}^{-1/2}$; 100-ms crosscorrelation length.}
\begin{tabular}{|c|c|c|c|c|c|c|c|c|c|c|c|c|c|c|c|c|c|c|c|}
\hline
\hline
                           & \multicolumn{3}{c|}{100 Hz} & \multicolumn{3}{c|}{150 Hz}  & \multicolumn{3}{c|}{250 Hz}  &
                             \multicolumn{3}{c|}{554 Hz} & \multicolumn{3}{c|}{1000 Hz} & \multicolumn{3}{c|}{1850 Hz} \\\cline{2-19}
GRB     &      &      &      &      &      &      &      &      &      &      &      &      &      &      &      &      &      &      \\
date & H1-H2 & H1-L1 & H2-L1 & H1-H2 & H1-L1 & H2-L1 &
       H1-H2 & H1-L1 & H2-L1 & H1-H2 & H1-L1 & H2-L1 &
       H1-H2 & H1-L1 & H2-L1 & H1-H2 & H1-L1 & H2-L1 \\
\hline
\hline
050223  &  5.6 &  ... &  ... &  4.1 &  ... &  ... &  4.8 &  ... &  ... &  8.0 &  ... &  ... & 14.5 &  ... &  ... & 30.9 &  ... &  ... \\
050306  &  6.9 &  6.7 & 12.6 &  4.9 &  5.8 &  9.1 &  5.6 &  7.6 & 10.4 &  9.1 & 13.8 & 17.3 & 16.0 & 28.0 & 34.0 & 30.0 & 74.1 & 91.8 \\
050318  &  7.4 &  9.7 & 12.5 &  5.9 &  7.4 & 10.3 &  6.4 &  9.9 & 11.8 & 10.7 & 17.5 & 17.9 & 18.4 & 33.2 & 34.1 & 33.3 & 63.4 & 64.5 \\
050319  &  5.5 &  6.0 &  9.6 &  4.6 &  4.6 &  7.2 &  5.2 &  6.5 &  8.4 &  8.8 & 11.4 & 14.4 & 15.2 & 21.3 & 25.1 & 30.1 & 34.7 & 48.3 \\
\hline
\end{tabular}

\caption{\label{tab:s3ul100ms}S3 90\% upper limits on hrss of Q = 8.9 linearly polarized sine-gaussians, in units of
$10^{-20}\:\text{Hz}^{-1/2}$; 100-ms crosscorrelation length.}
\begin{tabular}{|c|c|c|c|c|c|c|c|c|c|c|c|c|c|c|c|c|c|c|c|}
\hline
\hline
                           & \multicolumn{3}{c|}{100 Hz} & \multicolumn{3}{c|}{150 Hz}  & \multicolumn{3}{c|}{250 Hz}  &
                             \multicolumn{3}{c|}{554 Hz} & \multicolumn{3}{c|}{1000 Hz} & \multicolumn{3}{c|}{1850 Hz} \\\cline{2-19}
GRB     &      &      &      &      &      &      &      &      &      &      &      &      &      &      &      &      &      &      \\
date & H1-H2 & H1-L1 & H2-L1 & H1-H2 & H1-L1 & H2-L1 &
       H1-H2 & H1-L1 & H2-L1 & H1-H2 & H1-L1 & H2-L1 &
       H1-H2 & H1-L1 & H2-L1 & H1-H2 & H1-L1 & H2-L1 \\
\hline
\hline
031108  &  6.0 &  ... &  ... &  3.6 &  ... &  ... &  3.8 &  ... &  ... &  4.5 &  ... &  ... &  7.9 &  ... &  ... & 20.1 &  ... &  ... \\
031109a &  4.4 &  ... &  ... &  2.7 &  ... &  ... &  2.9 &  ... &  ... &  3.5 &  ... &  ... &  6.1 &  ... &  ... & 15.1 &  ... &  ... \\
031220  &  5.0 &  ... &  ... &  3.0 &  ... &  ... &  3.0 &  ... &  ... &  4.1 &  ... &  ... &  7.0 &  ... &  ... & 15.8 &  ... &  ... \\
\hline
\end{tabular}

\caption{\label{tab:s2ul100ms}S2 90\% upper limits on hrss of Q = 8.9 linearly polarized sine-gaussians, in units of
$10^{-19}\:\text{Hz}^{-1/2}$; 100-ms crosscorrelation length.}
\begin{tabular}{|c|c|c|c|c|c|c|c|c|c|c|c|c|c|c|c|c|c|c|c|}
\hline
\hline
                           & \multicolumn{3}{c|}{100 Hz} & \multicolumn{3}{c|}{150 Hz}  & \multicolumn{3}{c|}{250 Hz}  &
                             \multicolumn{3}{c|}{554 Hz} & \multicolumn{3}{c|}{1000 Hz} & \multicolumn{3}{c|}{1850 Hz} \\\cline{2-19}
GRB     &      &      &      &      &      &      &      &      &      &      &      &      &      &      &      &      &      &      \\
date & H1-H2 & H1-L1 & H2-L1 & H1-H2 & H1-L1 & H2-L1 &
       H1-H2 & H1-L1 & H2-L1 & H1-H2 & H1-L1 & H2-L1 &
       H1-H2 & H1-L1 & H2-L1 & H1-H2 & H1-L1 & H2-L1 \\
\hline
\hline
030217  &  ... &  ... &  4.0 &  ... &  ... &  2.0 &  ... &  ... & 0.94 &  ... &  ... &  1.5 &  ... &  ... &  4.1 &  ... &  ... &  9.5 \\
030226  &  7.3 &  3.1 &  5.3 &  3.2 &  1.5 &  2.1 &  1.1 & 0.65 & 0.62 &  1.4 &  1.0 & 0.85 &  2.6 &  2.4 &  1.4 &  7.1 &  6.5 &  2.7 \\
030320a &  6.7 &  2.3 &  6.8 &  2.5 &  1.3 &  2.3 & 0.76 & 0.67 & 0.70 &  1.0 &  1.2 &  1.4 &  1.6 &  3.6 &  3.5 &  4.1 &  7.2 &  5.8 \\
030323a &  5.3 &  2.7 &  5.6 &  3.0 &  1.5 &  2.5 &  1.2 & 0.86 &  1.4 &  1.8 &  2.2 &  3.0 &  2.7 &  5.5 &  7.0 &  6.4 & 10.0 & 12.4 \\
030323b &  5.1 &  1.8 &  4.9 &  2.0 & 0.95 &  1.7 & 0.77 & 0.47 & 0.79 &  1.1 & 0.90 &  1.6 &  1.6 &  1.9 &  2.5 &  3.9 &  3.7 &  5.0 \\
030324  &  8.7 &  ... &  ... &  4.6 &  ... &  ... &  1.5 &  ... &  ... &  2.0 &  ... &  ... &  3.7 &  ... &  ... &  8.0 &  ... &  ... \\
030325  &  2.9 &  1.5 &  3.4 &  1.4 & 0.78 &  1.6 & 0.63 & 0.46 & 0.90 &  1.0 & 1.00 &  1.9 &  1.5 &  1.9 &  2.9 &  3.7 &  4.6 &  6.6 \\
030326  &  9.0 &  3.0 &  7.4 &  4.2 &  1.8 &  3.1 &  1.3 & 0.81 & 0.98 &  1.9 &  1.5 &  1.8 &  3.7 &  3.1 &  2.9 &  8.6 &  6.8 &  5.7 \\
030329a &  4.4 &  ... &  ... &  2.5 &  ... &  ... &  1.2 &  ... &  ... &  2.1 &  ... &  ... &  3.0 &  ... &  ... &  8.6 &  ... &  ... \\
030329b &  2.6 &  ... &  ... &  1.2 &  ... &  ... & 0.34 &  ... &  ... & 0.56 &  ... &  ... & 0.94 &  ... &  ... &  2.2 &  ... &  ... \\
030331  &  ... &  3.5 &  ... &  ... &  1.7 &  ... &  ... & 0.97 &  ... &  ... &  2.1 &  ... &  ... &  4.1 &  ... &  ... & 10.3 &  ... \\
030405  &  2.3 &  1.2 &  2.6 &  1.3 & 0.76 &  1.1 & 0.46 & 0.40 & 0.47 & 0.73 & 0.73 & 0.90 &  1.2 &  1.8 &  1.9 &  2.7 &  4.4 &  4.0 \\
030406  &  ... &  1.2 &  ... &  ... & 0.73 &  ... &  ... & 0.45 &  ... &  ... & 0.87 &  ... &  ... &  1.9 &  ... &  ... &  5.0 &  ... \\
030413  &  ... &  ... &  1.7 &  ... &  ... & 0.94 &  ... &  ... & 0.61 &  ... &  ... &  1.1 &  ... &  ... &  2.9 &  ... &  ... &  5.4 \\
030414  &  1.3 &  ... &  ... & 0.89 &  ... &  ... & 0.30 &  ... &  ... & 0.43 &  ... &  ... & 0.74 &  ... &  ... &  1.7 &  ... &  ... \\
\hline
\end{tabular}
\end{table}
\end{turnpage}

\begin{turnpage}
\begin{table}
\caption{\label{tab:s4ul100mspsic}S4 90\% upper limits on hrss of Q = 8.9 circularly polarized sine-gaussians, in units of
$10^{-21}\:\text{Hz}^{-1/2}$; 100-ms crosscorrelation length.}
\begin{tabular}{|c|c|c|c|c|c|c|c|c|c|c|c|c|c|c|c|c|c|c|c|}
\hline
\hline
                           & \multicolumn{3}{c|}{100 Hz} & \multicolumn{3}{c|}{150 Hz}  & \multicolumn{3}{c|}{250 Hz}  &
                             \multicolumn{3}{c|}{554 Hz} & \multicolumn{3}{c|}{1000 Hz} & \multicolumn{3}{c|}{1850 Hz} \\\cline{2-19}
GRB     &      &      &      &      &      &      &      &      &      &      &      &      &      &      &      &      &      &      \\
date & H1-H2 & H1-L1 & H2-L1 & H1-H2 & H1-L1 & H2-L1 &
       H1-H2 & H1-L1 & H2-L1 & H1-H2 & H1-L1 & H2-L1 &
       H1-H2 & H1-L1 & H2-L1 & H1-H2 & H1-L1 & H2-L1 \\
\hline
\hline
050223  &  1.7 &  ... &  ... &  1.3 &  ... &  ... &  1.5 &  ... &  ... &  2.4 &  ... &  ... &  4.4 &  ... &  ... &  8.3 &  ... &  ... \\
050306  &  2.0 &  1.5 &  2.6 &  1.5 &  1.2 &  1.9 &  1.7 &  1.7 &  2.2 &  2.8 &  3.1 &  3.7 &  4.9 &  6.0 &  7.0 &  9.3 & 16.3 & 19.1 \\
050318  &  2.2 &  2.1 &  2.8 &  1.7 &  1.6 &  2.2 &  1.9 &  2.1 &  2.4 &  3.0 &  4.0 &  4.2 &  5.5 &  6.9 &  7.4 & 10.3 & 12.7 & 14.0 \\
050319  &  1.7 &  1.6 &  2.5 &  1.4 &  1.3 &  1.9 &  1.6 &  1.7 &  2.2 &  2.6 &  3.2 &  3.8 &  4.7 &  5.7 &  6.7 &  9.1 & 10.3 & 12.8 \\
\hline
\end{tabular}

\caption{\label{tab:s3ul100mspsic}S3 90\% upper limits on hrss of Q = 8.9 circularly polarized sine-gaussians, in units of
$10^{-21}\:\text{Hz}^{-1/2}$; 100-ms crosscorrelation length.}
\begin{tabular}{|c|c|c|c|c|c|c|c|c|c|c|c|c|c|c|c|c|c|c|c|}
\hline
\hline
                           & \multicolumn{3}{c|}{100 Hz} & \multicolumn{3}{c|}{150 Hz}  & \multicolumn{3}{c|}{250 Hz}  &
                             \multicolumn{3}{c|}{554 Hz} & \multicolumn{3}{c|}{1000 Hz} & \multicolumn{3}{c|}{1850 Hz} \\\cline{2-19}
GRB     &      &      &      &      &      &      &      &      &      &      &      &      &      &      &      &      &      &      \\
date & H1-H2 & H1-L1 & H2-L1 & H1-H2 & H1-L1 & H2-L1 &
       H1-H2 & H1-L1 & H2-L1 & H1-H2 & H1-L1 & H2-L1 &
       H1-H2 & H1-L1 & H2-L1 & H1-H2 & H1-L1 & H2-L1 \\
\hline
\hline
031108  & 18.4 &  ... &  ... & 11.5 &  ... &  ... & 11.8 &  ... &  ... & 14.0 &  ... &  ... & 23.2 &  ... &  ... & 61.0 &  ... &  ... \\
031109a & 13.5 &  ... &  ... &  8.5 &  ... &  ... &  8.7 &  ... &  ... & 11.3 &  ... &  ... & 19.0 &  ... &  ... & 47.6 &  ... &  ... \\
031220  & 12.1 &  ... &  ... &  9.4 &  ... &  ... &  8.8 &  ... &  ... & 11.6 &  ... &  ... & 20.5 &  ... &  ... & 49.1 &  ... &  ... \\
\hline
\end{tabular}

\caption{\label{tab:s2ul100mspsic}S2 90\% upper limits on hrss of Q = 8.9 circularly polarized sine-gaussians, in units of
$10^{-20}\:\text{Hz}^{-1/2}$; 100-ms crosscorrelation length.}
\begin{tabular}{|c|c|c|c|c|c|c|c|c|c|c|c|c|c|c|c|c|c|c|c|}
\hline
\hline
                           & \multicolumn{3}{c|}{100 Hz} & \multicolumn{3}{c|}{150 Hz}  & \multicolumn{3}{c|}{250 Hz}  &
                             \multicolumn{3}{c|}{554 Hz} & \multicolumn{3}{c|}{1000 Hz} & \multicolumn{3}{c|}{1850 Hz} \\\cline{2-19}
GRB     &      &      &      &      &      &      &      &      &      &      &      &      &      &      &      &      &      &      \\
date & H1-H2 & H1-L1 & H2-L1 & H1-H2 & H1-L1 & H2-L1 &
       H1-H2 & H1-L1 & H2-L1 & H1-H2 & H1-L1 & H2-L1 &
       H1-H2 & H1-L1 & H2-L1 & H1-H2 & H1-L1 & H2-L1 \\
\hline
\hline
030226  & 22.1 &  9.6 & 16.7 &  9.8 &  4.7 &  6.1 &  3.1 &  2.0 &  1.9 &  4.1 &  3.2 &  2.6 &  7.8 &  7.2 &  4.5 & 21.2 & 19.6 &  8.0 \\
030320a & 21.0 &  7.4 & 24.7 &  7.7 &  4.1 &  7.6 &  2.2 &  2.2 &  2.4 &  3.3 &  4.0 &  4.5 &  5.3 & 11.4 & 10.8 & 12.3 & 21.7 & 18.3 \\
030323a & 16.7 &  ... &  ... &  8.8 &  ... &  ... &  4.0 &  ... &  ... &  6.3 &  ... &  ... &  9.3 &  ... &  ... & 21.8 &  ... &  ... \\
030323b & 14.8 &  4.8 & 14.4 &  5.9 &  2.6 &  4.8 &  2.2 &  1.3 &  2.4 &  3.3 &  2.5 &  4.4 &  4.7 &  5.3 &  7.4 & 10.9 & 10.4 & 14.1 \\
030324  & 27.0 &  ... &  ... & 13.9 &  ... &  ... &  4.7 &  ... &  ... &  6.3 &  ... &  ... & 10.7 &  ... &  ... & 24.7 &  ... &  ... \\
030325  &  9.7 &  3.7 &  9.9 &  4.6 &  2.0 &  4.5 &  2.2 &  1.2 &  2.8 &  3.5 &  2.6 &  5.4 &  5.2 &  5.1 &  8.4 & 12.3 & 12.5 & 19.1 \\
030326  & 28.3 & 11.0 & 28.6 & 13.0 &  6.3 & 10.9 &  4.3 &  2.9 &  3.7 &  6.3 &  5.0 &  6.2 & 10.6 & 10.6 & 10.2 & 26.4 & 23.2 & 18.3 \\
030329a & 13.7 &  ... &  ... &  7.8 &  ... &  ... &  3.6 &  ... &  ... &  5.8 &  ... &  ... &  9.5 &  ... &  ... & 24.8 &  ... &  ... \\
030329b &  8.1 &  ... &  ... &  3.3 &  ... &  ... &  1.0 &  ... &  ... &  1.7 &  ... &  ... &  2.8 &  ... &  ... &  6.7 &  ... &  ... \\
030331  &  ... &  7.4 &  ... &  ... &  3.7 &  ... &  ... &  2.1 &  ... &  ... &  4.9 &  ... &  ... &  8.6 &  ... &  ... & 20.6 &  ... \\
030405  &  7.1 &  3.1 &  6.8 &  3.7 &  1.9 &  2.9 &  1.3 &  1.1 &  1.2 &  2.1 &  1.9 &  2.3 &  3.3 &  5.0 &  5.2 &  7.8 & 11.6 & 10.4 \\
030406  &  ... &  2.8 &  ... &  ... &  1.7 &  ... &  ... &  1.0 &  ... &  ... &  2.0 &  ... &  ... &  4.6 &  ... &  ... & 11.6 &  ... \\
030413  &  ... &  ... &  4.3 &  ... &  ... &  2.4 &  ... &  ... &  1.5 &  ... &  ... &  2.8 &  ... &  ... &  7.3 &  ... &  ... & 13.5 \\
030414  &  4.1 &  ... &  ... &  2.7 &  ... &  ... & 0.91 &  ... &  ... &  1.3 &  ... &  ... &  2.2 &  ... &  ... &  5.2 &  ... &  ... \\
\hline
\end{tabular}
\end{table}
\end{turnpage}

\end{document}